\documentclass[12pt]{article}
\usepackage{amsmath}
\usepackage{graphicx}
\usepackage{enumerate}

\usepackage{url} % not crucial - just used below for the URL 
\usepackage{tabularx} % extra features for tabular environment

\DeclareMathOperator*{\argmin}{arg\,min}
\usepackage{diagbox}

\usepackage{booktabs}
\usepackage{siunitx}
\usepackage[english]{babel}
\usepackage{amsthm}
\usepackage{amssymb}
\usepackage{multirow}
\usepackage{xr}
\usepackage{array,multirow}
\usepackage{booktabs}
\usepackage{subcaption}
\usepackage{mwe}
\usepackage[export]{adjustbox}
\usepackage{enumitem}
\usepackage{listings}
\usepackage{threeparttable}
\usepackage{romannum}
\usepackage{bm}
\usepackage{bbm}
\usepackage{rotating}
\usepackage[round]{natbib}
\newtheorem{theorem}{Theorem}[]
\newtheorem{corollary}{Corollary}[]

\usepackage[skip=1ex]{caption}

\usepackage[ruled,vlined]{algorithm2e}
\usepackage{xcolor}

\usepackage{float}
\usepackage[utf8]{inputenc}
\usepackage{textcomp}
\usepackage{tabularx,ragged2e,booktabs,caption}
\newcolumntype{C}[1]{>{\Centering}m{#1}}

\def\T{{ \intercal }}

\usepackage{float}

\newtheorem{assumption}{Assumption}
\newtheorem{proposition}{Proposition}
\newtheorem{remark}{Remark}

\newcommand{\wb}{\mathbf{w}}

\newcommand{\br}{\bm{r}}

\newcommand{\by}{\bm{y}}

\newcommand{\Ab}{\mathbf{A}}
\newcommand{\Bb}{\mathbf{B}}

\newcommand{\Fb}{\mathbf{F}}
\newcommand{\Gb}{\mathbf{G}}
\newcommand{\Hb}{\mathbf{H}}

\newcommand{\Ub}{\mathbf{U}}

\newcommand{\Xb}{\mathbf{X}}
\newcommand{\Yb}{\mathbf{Y}}

\newcommand{\bP}{\bm{P}}

\newcommand{\bU}{\bm{U}}

\newcommand{\bX}{\bm{X}}

\newcommand{\bZ}{\bm{Z}}

\newcommand{\cB}{\mathcal{B}}

\newcommand{\cL}{\mathcal{L}}

\newcommand{\cN}{\mathcal{N}}

\newcommand{\EE}{\mathbb{E}}

\newcommand{\RR}{\mathbb{R}}

\newcommand{\bbeta}{\bm{\beta}}
\newcommand{\bgamma}{\bm{\gamma}}

\newcommand{\bxi}{\bm{\xi}}

\newcommand{\bphi}{\bm{\phi}}
\newcommand{\bvarphi}{\bm{\varphi}}

\newcommand{\bGamma}{\bm{\Gamma}}

\newcommand{\bSigma}{\bm{\Sigma}}

\newcommand*{\zero}{{\bm 0}}
\newcommand*{\one}{{\bm 1}}

\def\T{{ \intercal }}

\makeatletter
\newcommand*{\addFileDependency}[1]{% argument=file name and extension
  \typeout{(#1)}
  \@addtofilelist{#1}
  \IfFileExists{#1}{}{\typeout{No file #1.}}
}
\makeatother

\newcommand*{\myexternaldocument}[1]{%
    \externaldocument{#1}%
    \addFileDependency{#1.tex}%
    \addFileDependency{#1.aux}%
}
\myexternaldocument{supp_K_grow}

\def\spacingset#1{\renewcommand{\baselinestretch}%
{#1}\small\normalsize} \spacingset{1}

\begin{document}
\pagenumbering{arabic}

\title{\Large Inference on Generalized Latent Variable Models with High-Dimensional Responses and Covariates %: A Scalable Alternating Approach
}
\date{}
\author{\normalsize Jing Ouyang$^1$, Chengyu Cui$^2$, Yunxiao Chen$^3$, Kean Ming Tan$^2$, and Gongjun Xu$^2$\\~\\
\small 1. Faculty of Business and Economics, University of Hong Kong\\
\small 2. Department of Statistics, University of Michigan\\
\small 3. Department of Statistics, London School of Economics and Political Science}
%\author{}
\maketitle

 \spacingset{1.75} % DON'T change the spacing

\begin{abstract}

Regression models with both high-dimensional responses and covariates have attracted growing attention. Standard multivariate regression models become inadequate when the response variables depend not only on observed covariates but also on latent variables that capture key unobserved characteristics. To draw statistical inferences on covariate effects while accounting for latent variables, we consider a high-dimensional generalized latent variable model that accommodates mixed-type responses and allows for flexible dependence between covariates and latent variables, which is more suitable for many real-world applications than existing methods that either rely on a linear regression form or restricted assumptions on the dependence between covariates and latent variables. 
We develop an alternating algorithm that iteratively updates the regression parameters and the latent variables, transforming an intractable nonconvex problem into a sequence of tractable convex subproblems. Theoretically, we provide algorithmic guarantees by establishing statistical consistency of the resulting estimator and deriving an error bound for it. Further, building on this estimator, we construct a debiased estimator for the covariate effect and establish its asymptotic normality. The effectiveness of the proposed method is demonstrated through an application to evaluating the fairness of the Programme for International Student Assessment (PISA).

\noindent {\em Keywords:} Nonlinear Factor Models; Latent Confounders; Regularized Estimation; Nonconvex Optimization; Alternating Algorithm

\end{abstract}

%\newpage

\section{Introduction}

% \yc{Suggested flow: 
% \begin{enumerate}
%     \item High-dimensional regression: high-dimensional X. Then multivariate regression: both high-dimensional X and Y. 

%     \item High-dimensional multivariate regression with latent variables. The existing models. Applications. And then the limitations in these models. Maybe mention the literature of latent confounders (e.g., Wang Yixing's work? I am not very sure about this).

%     \item The current proposal and how it differs from the existing ones (with the table). Explain its challenges and how we deal with these challenges. 
% \end{enumerate}
% }

Regression with high-dimensional covariates, where the covariate dimension is of the same order as, or larger than, the sample size, is widely studied and has broad applications in education, political science, economics, and related fields~\citep{buhlmann2011statistics,wainwright2019high}. 
In the univariate-response setting, extensive work establishes estimation consistency and valid inference for low-dimensional components of high-dimensional regression parameters under sparsity assumptions~\citep{zhang2014,van2014,javanmard2014confidence}. 
However, in many modern applications, one observes a large number of responses for each subject, leading to regression problems with multivariate responses. This motivates a growing body of work on estimation and inference in high-dimensional multivariate regression settings~\citep{xia2018joint, bing2019adaptive}.
Nonetheless, most of the existing literature focuses on models in which all variables are fully observed.

However, in many scientific applications, responses depend not only on high-dimensional covariates but also on latent variables that capture unobserved subject characteristics. 
A broad literature develops methods to adjust for latent variables and enable valid estimation and inference on covariate effects. 
Early contributions largely focus on linear frameworks that relate responses to low-dimensional covariates and latent variables~\citep[e.g.,][]{leek2008general, gagnon2012using, wang2017confounder, mckennan2019accounting}. 
More recently, motivated by modern applications with rich covariate information, researchers extend these ideas to settings with high-dimensional covariates and latent variables, and develop inferential procedures based on factor-adjustment or deconfounding techniques~\citep[e.g.,][]{guo2021doubly, bing2023inference, fan2024latent}. 
Moving beyond linear models, an increasing amount of work investigates generalized latent variable models, including methods developed for settings with low-dimensional covariates and latent variables~\citep{du2025simultaneous, ouyang2025statistical} and approaches that accommodate high-dimensional covariates and latent variables~\citep{ouyang2023high, wang2025latent, lee2025ghive}.
These latent variable frameworks arise in various applications.
In educational testing, students' item responses are driven by latent abilities and a high-dimensional set of demographic and socioeconomic covariates that describe students' learning and living environments. 
Adjusting for latent abilities is essential for inferring covariate effects when evaluating testing fairness, that is, whether students from different demographic backgrounds have equal probabilities endorsing the test item, after controlling for their latent proficiency~\citep{chen2023dif,ouyang2025statistical}. 
In genomics and biomedical studies, latent biological factors (e.g., batch effects or population structures) may confound the relationship between high-dimensional gene expression status and health outcomes; accounting for latent variables is crucial for valid inference on covariate effects~\citep{leek2008general, gagnon2012using, wang2017confounder, ouyang2023high, du2025simultaneous}. 
In political science, voting behavior is often analyzed using high-dimensional regression models with rich demographic covariates, such as income, ethnicity, age, region, and their interactions~\citep{pandolfi2024conjugate, goplerud2025partially}. In this setting, incorporating latent political ideologies as latent variables can further improve model fit and provide deeper insight into the underlying structure that drives political behavior.

Despite these methodological advances and broad applications, existing methods often face major limitations. 
Many methods focus on a linear framework relating responses to covariates and latent variables and, therefore, are not applicable to accommodate mixed-type responses~\citep{leek2008general, gagnon2012using, wang2017confounder, lee2017improved, mckennan2019accounting, Domagoj2020, guo2021doubly, bing2023inference, fan2024latent}. Some recent approaches under a nonlinear framework impose restrictive dimensionality assumptions that cannot handle high-dimensional covariates~\citep{ du2025simultaneous, ouyang2025statistical}
or rely on a dependence structure, such as a linear factor model to relate covariates and latent variables~\citep{ouyang2023high, wang2025latent, lee2025ghive}. See Table~\ref{tab:selective works} for a detailed comparison of relevant approaches.
Overall, it remains an important open problem to develop  inference procedures in a nonlinear regression framework, accommodating high-dimensional responses and covariates, and allowing flexible dependence between covariates and latent structures.
Moreover, the existing literature provides a limited understanding of the algorithmic properties of the proposed estimation procedures. Even for generalized latent variable models without covariates, establishing algorithmic guarantees, such as error bounds of algorithm updates, remains an open problem.

\begin{table}
\centering
\resizebox{0.95\textwidth}{!}{
\begin{tabular}{cccccc}
\toprule
 &\begin{tabular}{c} Linear \\ Framework \end{tabular} &  \begin{tabular}{c} Nonlinear \\ Framework \end{tabular} & \begin{tabular}{c} High-Dimensional \\ Covariates \end{tabular} & \begin{tabular}{c} Flexible Correlation between \\
 Covariates and Latent Variables \end{tabular} & \begin{tabular}{c} Algorithmic \\ Guarantee\end{tabular} \\\midrule
 \begin{tabular}{c}
  %  SVA-related methods:\\
    \cite{leek2008general} \\
     \cite{gagnon2012using} \\
     \cite{wang2017confounder} \\
     \cite{lee2017improved} \\
     \cite{mckennan2019accounting}
     
\end{tabular}  & {\Huge $\checkmark$} &  {\Huge $\times$}  & {\Huge $\times$}  & {\Huge $\times$}  & {\Huge $\times$} \\\cmidrule(lr){1-6}
\begin{tabular}{c}
\cite{Domagoj2020} \\
     \cite{guo2021doubly} \\
     \cite{fan2024latent}  \end{tabular} & {\Huge $\checkmark$} & {\Huge $\times$} & {\Huge $\checkmark$} & {\Huge $\times$} & {\Huge $\times$}\\\cmidrule(lr){1-6}
  \begin{tabular}{c}
     \cite{bing2023inference}  \end{tabular} & {\Huge $\checkmark$} & {\Huge $\times$} & {\Huge $\checkmark$} & {\Huge $\checkmark$} & {\Huge $\times$}\\\cmidrule(lr){1-6}   
\begin{tabular}{c}
\cite{ouyang2023high} \\
\cite{wang2025latent} \\
\cite{lee2025ghive}
\end{tabular} & {\Huge $\checkmark$}& {\Huge $\checkmark$}  & {\Huge $\checkmark$} & {\Huge $\times$}  & {\Huge $\times$}\\\cmidrule(lr){1-6}
\begin{tabular}{c}\cite{du2025simultaneous}\\
\cite{ouyang2025statistical} \end{tabular}  & {\Huge $\checkmark$}&  {\Huge $\checkmark$}& {\Huge $\times$}  & {\Huge $\checkmark$} & {\Huge $\times$} \\\cmidrule(lr){1-6}
{\bf\large This paper} & {\Huge $\checkmark$} & {\Huge $\checkmark$} & {\Huge $\checkmark$} & {\Huge $\checkmark$} & {\Huge $\checkmark$} \\\bottomrule
\end{tabular}}
\caption{Comparison of our work with relevant approaches.}
\label{tab:selective works}
\end{table}

Establishing inference theory in such a general setting is intrinsically challenging. For regression with high-dimensional covariates, it is common to assume sparsity in the covariate effects and to estimate the regression coefficients using regularization methods that encourage sparse solutions~\citep{wainwright2019high}. However, in the presence of latent variables, regularized estimation poses substantial challenges for statistical inference, as the resulting optimization problem is highly nonconvex and the latent variables are intricately coupled with the high-dimensional covariates.
In linear models, this difficulty can sometimes be mitigated by projection or decomposition techniques that make use of the linear additivity and explicit residualization forms to separate regression parameters from latent variables. 
Within a nonlinear framework that relates responses to high-dimensional covariates and latent variables, these techniques are no longer applicable; see a detailed discussion of challenges in Section~\ref{sec:challenges}.

To address these challenges, we develop an alternating algorithm that efficiently estimates the model parameters and enables valid inference on covariate effects.
This algorithm converts an intractable nonconvex optimization problem into a sequence of tractable convex subproblems by alternating between updating the covariate effects and updating latent variables, each of which can be implemented efficiently using gradient-based algorithms.
Beyond computation, a major contribution of this work is a theoretical characterization of the algorithm itself. We address this open 
problem in the literature by establishing error bounds for estimators obtained from the alternating algorithms under generalized latent variable models. 
Building on these consistent estimators, we further construct a debiased estimator for the covariate effect of interest, using the estimators from alternating algorithms as initialization. 
We establish asymptotic normality results for the debiased estimator, enabling valid inference on the covariate effect in the presence of latent variables and nonlinear link functions.

The rest of the paper is organized as follows. Section~\ref{sec:model} introduces the model setup and elaborates on key challenges of estimation and inference. Section~\ref{sec:method} presents the alternating algorithm and the subsequent inference procedure. Section~\ref{sec:theory} establishes the estimation consistency for all model parameter estimators and the statistical inference for covariate effects. In Section~\ref{sec:simulation}, we perform extensive simulation studies to demonstrate the effectiveness of our method and the validity of inference theory. In Section~\ref{sec:data application}, we apply the proposed method to analyze real-world data, the Programme for International Student Assessment (PISA) 2022, and provide meaningful scientific findings. Section~\ref{sec:discussion} concludes this paper with discussions and potential directions for future research. Additional numerical studies and proofs are provided in the Supplementary Material.
The code for simulation studies and real data application is available at \url{https://anonymous.4open.science/r/High-Dim-GLVM-E44C}.

\noindent{\em Notation.} For any integer $N$, let $[N] = \{1, \dots, N\}$. For any set $S$, let $\mathrm{card}(S)$ denote its cardinality.
For any vector $\br = (r_1, \dots, r_l)^{\T}$, let $\| \br\|_0  = \text{card} (\{j: r_j \neq 0 \})$,  $\|\br \|_{\infty}= \max_{j = 1, \ldots, l} |r_j|$, and $\|\br\|_q = (\sum_{j=1}^l |r_j|^q)^{1/q}$ for $q \geq 1$.
For a matrix $\Ab = (a_{ij})_{n\times l}$, let $\|\Ab\|_{\infty,1} = \max_{j=1,\ldots, l} $ $\sum_{i = 1}^n |a_{ij} |$ to be the maximum absolute column sum, $\| \Ab\|_{1,\infty} = \max_{i=1,\ldots, n} \sum_{j=1}^l |a_{ij}| $ to be the maximum of the absolute row sum, $\| \Ab\|_{\max} = \max_{i,j} |a_{ij}|$ to be the maximum of the matrix entry,  and $\| \Ab\|_{F} = (\sum_{i=1}^n \sum_{j=1}^l |a_{ij}|^2)^{1/2}$ to be the Frobenius norm of $\Ab$. For any square matrix $\Ab=(a_{ij})_{n\times n}$, let $\lambda_{\min}(\Ab)$ and $\lambda_{\max}(\Ab)$ be the smallest and largest eigenvalues of $\Ab$, respectively. For sequences $\{a_n\}$ and $\{b_n\}$, we write $a_n \lesssim b_n $ if there exists a constant $C >0$ such that $a_n \leq C b_n$ for all $n$, and $a_n \asymp b_n$ if $a_n \lesssim b_n $ and $b_n \lesssim a_n$.  %Let $\| \cdot \|_{\varphi_1}$ be sub-exponential norm and $\| \cdot \|_{\varphi_2}$ be sub-Gaussian norm. 

\section{Model Setup}	

%[Need to add discussion on $K$ is pre-specified, the number of latent variables is known. Confirmatory structure on $\bgamma$. Starting point is ``non convex'']

\label{sec:model}
\subsection{Generalized Latent Variable Models}
\label{sec:GLVM model}
Consider a setting where $n$ subjects respond to $q$ items. For subject $i$, we let $\bX_i \in \RR^p$ denote the observed covariates, such as language of assessments in educational testing or demographic characteristics in political science, where the covariate dimension $p$ can be of much higher order than $n$. We further let $\bm{U}_i \in \RR^K$ denote its latent variables, such as latent skills in educational testing or latent political ideology in political science. The subject's responses to the $q$ items are collected into a $q$-dimensional response vector $\by_i=(y_{i1},\dots,y_{iq})^\T$. 
For $j \in[q]$, we assume the following conditional distribution of $y_{ij}$:
\begin{equation}
	y_{ij}|w_{ij} \sim p_{ij}(\cdot \mid w_{ij}), \quad \text{ where } w_{ij} = \beta_{j0} + \bgamma_j^{\intercal}\bU_i  + \bbeta_j^{\intercal}\bX_i .\label{eq:model}
\end{equation}
In this model formulation, we let $\beta_{j0} \in \RR $ be the intercept and $\bbeta_j \in \RR^p$ be the coefficient parameters for the observed covariates, and let $\bgamma_j \in \RR^K$ be the loading parameters. 
 The function $p_{ij}(\cdot \mid w_{ij})$ is a probability density (or mass) function with a known parametric form that may vary across $i$ and $j$ and is typically specified according to the data types of the responses $y_{ij}$. %The conditional probability function is determined by $w_{ij}$ and is allowed to vary across $i$ and $j$, accommodating mixed data types for responses $y_{ij}$. 
For notational convenience, we use $\bphi = (\bbeta_0, \Bb, \bGamma, \Ub)$ to define the collection of all parameters throughout the manuscript, where $\Bb = (\bbeta_1, \dots, \bbeta_q)_{q \times p}^{\intercal}$ is the collection of covariate effect across all items, $\bGamma = (\bgamma_1, \dots, \bgamma_q)_{q\times K}^{\intercal}$ aggregates loading vectors for all items and $\Ub = (\bU_1, \dots, \bU_n)_{n \times K}^{\intercal}$ aggregates latent variables for all subjects.

High-dimensional generalized latent variable models arise in a wide range of applications. 
In educational assessments, students' responses $\by_i$ depend on latent abilities $\bU_i$ and a large number of observed covariates $\bX_i$, such as demographics (e.g., gender and ethnicity) and information on students' learning environments~\citep{holland2012differential}. 
The covariate effects $\bbeta_j$ quantify how the covariates affect the response to item $j$, conditioned on latent abilities. 
A similar structure appears in political science, where voting outcomes $\by_i$ may depend on many observed variables $\bX_i$ (e.g., income, ethnicity, and their interactions) while also being affected by a low-dimensional latent political ideology $\bU_i$~\citep{pandolfi2024conjugate, goplerud2025partially}. In this setting, $\bbeta_j$
 captures how the observed variables are associated with voting outcomes, conditioned on the latent ideology.

In modern large-scale educational assessments, practitioners typically specify the number of latent abilities $K$ a priori as part of the assessment framework or test design~\citep{schleicher2019pisa, oecd2024}. Similarly, in political science, latent ideologies are commonly modeled as low-dimensional and prespecified.
This assumption is also standard in confirmatory factor analysis~\citep{skrondal2004generalized,bartholomew2011latent}.
Motivated by these applications, we assume that the number of latent variables $K$ is known throughout the paper.
%\yc{It may read better if you move the discussion about the K being known to the previous paragraph. Then, the writing ``motivated by these applications" makes sense.}
 When $K$ is unknown, 
it can be selected using statistical procedures such as information-criterion methods~\citep{bai2002determining, chen2022determining} and parallel analysis~\citep{dobriban2020permutation}.

%** check the cited ref

%** Irini book, Anderse + sophia latent variable model book

%[change to political science example]
%A similar structure appears in genetics, where gene expression $\by_i$ may depend on a large number of experimental conditions and biological annotations recorded in $\bX_i$. At the same time, expression levels are influenced by latent biological variables such as batch effects or cell-type heterogeneity~\citep{wang2017confounder, ouyang2023high}. These settings also require modeling high-dimensional covariate effects $\bbeta_j$ incorporating latent variables.

%[merge into above applications]

\subsection{Background and Challenges}
\label{sec:challenges}

%\yc{Maybe just use the section title ``Background and Challenges"}
Estimation under the generalized latent variable model~\eqref{eq:model} and valid inference for the covariate effect are intrinsically challenging. 
To address these challenges, prior research has examined related settings, yielding a range of estimation and inference methods.
For linear models, a common approach is to assume sparsity on the covariate effects and estimate them by optimizing an $L_1$-regularized objective function~\citep{loh2013regularized,van2014,zhang2014,ning2017}. 
Specifically, the objective is formulated as the sum of a squared-loss function and an $L_1$ regularization term:
\begin{align}
    -\frac{1}{nq} \sum_{i=1}^n \sum_{j=1}^q\{ Y_{ij} - ( \beta_{j0} + \bbeta_j^{\T}\bX_i + \bgamma_j^{\T} {\bU}_i )\}^2 +\lambda \sum_{j=1}^q\|\bbeta_j\|_1.\label{eq:joint optimization}
\end{align}
Even in this linear setting, directly minimizing~\eqref{eq:joint optimization} in one step and establishing valid inference for the regularized estimator is highly nontrivial, because the objective is inherently nonconvex due to the tight coupling of the sparse regression coefficient $\bbeta_j$ and the latent variables $\bU_i$. To address this difficulty,
\cite{Domagoj2020}, \cite{guo2021doubly}, and \cite{fan2024latent} establish consistency and inference results by additionally assuming a latent factor structure to relate covariates $\bX_i$ and latent variables $\bU_i$.
This structure enables the separation of latent variables from the covariates. 
The adjustment is carried out either through a spectral transformation that shrinks the leading singular directions of the design matrix, or by first estimating the latent factor space from $\Xb$ and then applying projection or factor-adjusted debiasing procedures. 
By contrast,~\cite{bing2023inference} does not impose such a structural assumption, but still relies on the linearity of the model to project the responses onto the orthogonal complement of the estimated latent space to adjust the hidden effect.
%\textcolor{red}{This Bing et al sentence seems disconnected from the previous sentences. Add a sentence or some words to bridge?  }
 
 %extending these methods to generalized latent variable models is challenging. When the squared-loss term in~\eqref{eq:joint optimization} is replaced by a general loss, the latent effects enter through a nonlinear index and the score/Hessian depend on the unknown linear predictors. As a result, the closed-form residualization and decomposition steps that projections no longer preserve the likelihood or cleanly separate covariate effects from latent structure. The sparse and low-rank components remain tightly coupled, which substantially complicates both one-shot regularized estimation and the subsequent inference theory.

% [1. elaborate on linear case, how these decomposition techniques work, and how the assumption on the structure of $\bX_i$ and $\bU_i$ being used in these works   2. a special case in~\cite{bing2023inference} ]

% [Think about the sparsity level, are we paying more - need it to be even more sparse - to achieve flexible dependence?]

%[briefly mention, that glm is under the assumptions of extending linear models with additional assumptions, ]

Researchers have extended the existing approaches to nonlinear settings by considering a general regularized estimation problem, where the squared-loss term in~\eqref{eq:joint optimization} is replaced by a general loss
\begin{align}
    - \frac{1}{nq} \sum_{i=1}^n \sum_{j=1}^ql_{ij} ( \beta_{j0} + \bbeta_j^{\T}\bX_i + \bgamma_j^{\T} {\bU}_i ) +\lambda \sum_{j=1}^q\|\bbeta_j\|_1.\label{eq:joint optimization nonlinear}
\end{align}
The function $l_{ij} (w_{ij}) = \log p_{ij} (y_{ij} | w_{ij}) $ is the individual log-likelihood function under the generalized latent variable model \eqref{eq:model}.
Some existing works address the problem by imposing similar structural assumptions, such as assuming a linear factor model relating $\bX_i$ and $\bU_i$~\citep{ouyang2023high, wang2025latent} or by imposing certain separation conditions such as $\bP_{\Gamma} \Bb = \bm{0}$, where  $\bP_{\Gamma} = \bGamma (\bGamma^{\T} \bGamma)^{-1}\bGamma^{\T}$ denotes the projection matrix onto the column space of $\bGamma$~\citep{du2025simultaneous,lee2025ghive}.
%\yc{These two vectors are not of the same dimension. It is not very clear.}
%that enforce uncorrelatedness between $\bX_i$ and $\bU_i$
 These constraints mitigate the intrinsic difficulty by effectively separating covariate effects from latent variables. However, under more flexible dependence between $\bX_i$ and $\bU_i$ without these restrictions, establishing estimation and inference becomes substantially more challenging, if not infeasible. 
At a high level, the difficulty lies in separating covariate effects from latent structure; this challenge can be understood from two perspectives:

% Recent work has investigated estimation and inference in generalized latent variable models, by imposing additional structure, for example by assuming a linear factor model relating $\bX_i$ and $\bU_i$~\citep{ouyang2023high, wang2025latent} or by imposing orthogonality assumptions that enforce uncorrelatedness between $\bX_i$ and $\bU_i$~\citep{du2025simultaneous,lee2025ghive}. These constraints mitigate the intrinsic difficulty by effectively separating covariate effects from latent variables. However, under flexible dependence between $\bX_i$ and $\bU_i$ without these restrictions, establishing estimation and inference becomes substantially more challenging, if not infeasible. At a high level, the difficulty come from the inability to disentangle covariate effects from latent structure; this challenge can be understood from two perspectives:

\begin{itemize}
    \item From a geometric perspective, the high-dimensional estimation and inference are typically analyzed on a cone~\citep{wainwright2019high}. The estimation error of $\bbeta_j$ lies in the cone, and restricted strong convexity (RSC) is established within the cone space~\citep{loh2013regularized}. However, in our optimization problem~\eqref{eq:joint optimization}, the relevant cone jointly involves the sparse regression component $\bbeta_j$ and the loading parameter $\bgamma_j$. As a result, the curvature is analyzed jointly in these two directions, and the RSC is established in the joint direction of $(\bbeta_j, \bgamma_j)$, rather than in the direction of $\bbeta_j$ alone. 
In this case, there exist cone directions where estimation errors of $\bbeta_j$ generate small or even no curvature because they can be offset by estimation errors in $\bgamma_j$. This mixed cone structure makes it difficult to prove RSC and, consequently, to construct debiased estimators.

\item Algebraically, these difficulties are reflected in the Hessian's structure. Expanding the log-likelihood around the true parameter $\bphi^*$, we can observe that the Hessian is a high-dimensional block matrix with substantial off-diagonal terms, causing it to be non-invertible. Specifically, write $\bvarphi_j = (\beta_{j0}, \bgamma_j^{\T},\bbeta_j^{\T})^{\T}$ and $\bZ_i = (\bU_i^{\T},\bX_i^{\T})^{\T}$, the Hessian has a block form with $\Hb_{\varphi\varphi}(\bphi)$ denoting the second derivatives of general loss in~\eqref{eq:joint optimization nonlinear} with respect to $\bvarphi_j$’s respectively. 
For high-dimensional estimation and inference with known latent variables $\bU_i^*$, the $\Hb_{\varphi\varphi}(\bphi)$ is stably invertible~\citep{zhang2014, van2014, ning2017}. In our nonconvex setting, however, the sparse component $\bbeta_j$ and the latent variables $\bU_i$ are strongly coupled: the off-diagonal blocks $\Hb_{\varphi u}(\bphi)$ and $\Hb_{u\varphi}(\bphi)$, capturing the correlation between $\bvarphi_j$ and $\bU_i$, are of the same order as $\Hb_{\varphi\varphi}(\bphi)$. As a consequence, the Hessian is not stably invertible, and RSC cannot be established.

\end{itemize}

To address these challenges, we propose an alternating optimization approach in Section~\ref{sec:method}. %\textcolor{red}{Where is the method? Maybe say In section 3?} 
Theoretical justification for this method is provided in Section~\ref{sec:theory}.

\begin{remark}
A well-known challenge in generalized latent variable models is the rotational indeterminacy~\citep{skrondal2004generalized}. Specifically, the response distribution of $Y_{ij}$ remains unchanged under the transformation
\begin{align*}
    \{\bgamma_j, \bU_i \} \rightarrow \{ \tilde\bgamma_j, \tilde\bU_i\} = \{\Gb^{-\T} \bgamma_j, \Gb \bU_i\}, 
\end{align*}
for any invertible matrix $\Gb$, while leaving $\bbeta_j$ and $\beta_{j0}$ unchanged. 
Our inferential target is the covariate effect matrix $\Bb$, which does not require identifying $\bgamma_j$'s and $\bU_i$'s individually. In this work, we are primarily interested in drawing statistical inferences about the regression coefficients, and thus we do not impose additional constraints to identify the loadings or latent variables.
\end{remark}

	\section{Proposed Algorithm and Inference Method}
	\label{sec:method}
% {\color{red}
% you did not really write donw the likelihood function of model 1 yet?  For (3), $l_{ij}$ is not defined. $L$ in section 2 is also not defined. 
% This section needs to be written to emphasize the contribution more.  To address the challanges XXX, we propose an alternating optimization method that transforms XXX. S 3.1 directly starts with the algorithm, but reader has not even seen your joint optimization problem. It probably needs to be defined somewhere earlier? Instead of writing (2) perhaps you can directly introduce your joint optimization problem, and then when you talk about linear function, you mentioned identity function reduces to linear regression and it is also challenging?  }  

Motivated by the challenges of the joint optimization problem~\eqref{eq:joint optimization nonlinear}, we adopt an alternating optimization approach that converts solving a challenging nonconvex problem in one step into solving a sequence of tractable convex subproblems iteratively. Building on estimators from the alternating algorithm, we further construct a debiased estimator for covariate effects of interest, enabling valid statistical inference.

 \subsection{Alternating Algorithm}
\label{sec:alternating algorithm}

% The initialization of our algorithm is general and flexible. In particular, we will show that any consistent estimators for latent variables that satisfy mild rate conditions can serve as valid initializers. Details on the criteria for initial estimators are given in Section~\ref{sec:theory}, and initialization examples are discussed later in this section.

 % \textcolor{red}{U is not defined yet, no optimization problem.  Display equation you can use 1/n fractional form, make it easier to read.  }

Our algorithm starts with a suitable initial estimator of the latent variables, denoted by $\hat\bU_i^{(0)}$. The choice of initialization method can be case-specific, and several examples are provided later in this subsection. Given a valid initialization $\hat\Ub^{(0)}$, our alternating algorithm proceeds as follows.
  For each iteration $t = 1, 2, ...$, we treat the latent variables from the previous iteration, namely, $\hat{\bU}_i^{(t-1)}$'s as surrogate variables~\citep{leek2008general, wang2017confounder} and fit high-dimensional generalized linear models. Specifically, for $j \in [q]$, we update $\hat{\beta}_{j0}^{(t)}$, $\hat\bbeta_j^{(t)}$, and $\hat\bgamma_j^{(t)}$ by solving the $L_1$-regularized problem:
\begin{align}
     (\hat{\beta}_{j0}^{(t)}, \hat\bbeta_j^{(t)}, \hat\bgamma_j^{(t)}) &= 
     \underset{\beta_{j0}, \bbeta_j, \bgamma_j}
     {\argmin}~ - \frac{1}{n} \sum_{i=1}^n  l_{ij}( \beta_{j0} + \bbeta_j^{\T}\bX_i + \bgamma_j^{\T} \hat{\bU}_i^{(t-1)} )  +\lambda^{(t)}\|\bbeta_j\|_1. \label{eq:AM1}
\end{align}
We then plug the estimators $\hat{\bbeta}_{0}^{(t)}$, $\hat\bbeta_j^{(t)}$ and $\hat\bgamma_j^{(t)}$ into the original model and update the estimates for the latent variables. Specifically, for $i \in [n]$, we obtain $\hat{\bU}_i^{(t)}$ by solving the optimization problem: 
       \begin{align}
    		 \hat{\bU}_i^{(t)} &= \underset{\bU_i}{\argmin}~\ - \frac{1}{q} \sum_{j=1}^q l_{ij}( \hat\beta_{j0}^{(t)} + \hat\bgamma_j^{(t)\intercal}\bU_i + (\hat\bbeta_j^{(t)})^\T\bX_i). \label{eq:AM2}
	\end{align}

%{\color{blue} Only use $\hat{\bbeta}^{(t)}$, while we simulataneously update $\bU$ and $\bgamma$, and aim to achieve a better $\varepsilon_{nq}^{(t-1)}$ }

We iterate~\eqref{eq:AM1}--\eqref{eq:AM2} to update the item parameters $\hat{\beta}_{j0}^{(t)}, \hat\bbeta_j^{(t)}, \hat\bgamma_j^{(t)}$, and the latent variables $\hat{\bU}_i^{(t)}$.
Note that we do not require exact solutions to the problems \eqref{eq:AM1} and \eqref{eq:AM2} at each iteration $t$. Approximate first-order methods (e.g., gradient descent type updates; see Algorithms~1--2 in Supplementary Materials) are sufficient. %, provided that the algorithm error is of smaller order than the statistical error. 
In Proposition~\ref{prop:AM1 num steps} of Section~\ref{sec:theory}, we show that if we take $ M_1 \ge C_1\log n$ inner gradient steps for~\eqref{eq:AM1} and $M_2 \ge C_2 (\log n +\log q )$ inner gradient steps for \eqref{eq:AM2}, for sufficiently large constants $C_1$ and $C_2$, the resulting approximate estimators achieve the same convergence rates as that of the exact minimizers of~\eqref{eq:AM1} and~\eqref{eq:AM2}. Empirically, we iterate~\eqref{eq:AM1}--\eqref{eq:AM2} for multiple times and terminate the algorithm when $\max\{\|\hat{\Bb}^{(t)} - \hat{\Bb}^{(t-1)}\|_F, \|\hat{\bGamma}^{(t)} - \hat{\bGamma}^{(t-1)}\|_F, \|\hat{\Ub}^{(t)} - \hat{\Ub}^{(t-1)}\|_F, \|\hat{\bbeta}_0^{(t)} - \hat{\bbeta}_0^{(t-1)}\|_2\}$ is smaller than a pre-specified tolerance value or when a certain maximum iteration number is reached.

This alternating algorithm accommodates a broad class of initial estimators for the latent variables, as long as the initializer satisfies the mild rate conditions stated in Assumption~\ref{assumption:intialization} of Section~\ref{sec:theory}. Here we describe simple and scalable examples used in practice.
     One example of initialization is to set $\hat{\Bb}^{(0)} = \bm{0}$ and obtain the estimator $\hat\bGamma^{(0)}$ and $\hat\Ub^{(0)}$ by fitting a covariate-free working model, i.e., for sufficiently large constant $D$, we solve
\begin{align}
  & (\hat\bbeta_0^{(0)}, \hat\bGamma^{(0)}, \hat\Ub^{(0)}) = \underset{\max \{\|\bbeta_0\|_{\infty},  \|\bGamma\|_{\max}, \|\Ub\|_{\max}\} \le D }{\argmin} - \frac{1}{nq} \sum_{i=1}^n \sum_{j=1}^q l_{ij} ( \beta_{j0}+ \bgamma_j^{\intercal}\bU_i ).  
  \label{eq:initial estimation}
\end{align} 
Intuitively, \eqref{eq:initial estimation} corresponds to the maximum likelihood estimator of a covariate-free working model obtained by temporarily ignoring $\bX_i$. This provides a low-rank approximation to the response and is often a reasonable starting point when the latent variables capture the majority dependence pattern in the response and the covariate effects are relatively sparse. 

% \textcolor{red}{Why write the following via a remark? Seems like it is a discussion of the previous paragraph? Why not just discuss it without remark.  }
\begin{remark}
To compute \eqref{eq:initial estimation} efficiently, we adopt the SVD-based spectral initializer for exploratory factor analysis proposed by~\cite{zhang2020note}. In particular, we can apply Algorithm~1 of~\cite{zhang2020note} to the response matrix $\Yb$ under the covariate-free working model, which gives spectral estimates $(\hat\bbeta_0^{(0)},\hat\bGamma^{(0)},\hat\Ub^{(0)})$. This initializer is computationally efficient and scalable, making it a practical choice for the alternating algorithm.

In addition, we could obtain an initialization by replacing the explicit low-rank constraint in~\eqref{eq:initial estimation} with a nuclear-norm constraint, a common convex relaxation for low-rank component. This leads to a convex optimization problem that is typically straightforward to solve with off-the-shelf algorithms and comes with well-developed statistical guarantees for noisy low-rank estimation~\citep{koltchinskii2011nuclear, koltchinskii2015optimal}.

Finally, prior information can also be used for initialization. For instance, if a set of anchor items is known to have $\bbeta_j=\zero$, then these items can be used to obtain a more stable initial estimate of the latent variables. In educational applications, many items are reused and well-calibrated across PISA cycles and are typically treated as anchor items~\citep{oecd2024}. 
From these anchor items, we first estimate $(\bbeta_0,\bGamma,\Ub)$ under the covariate-free model. Then, given $\hat\Ub^{(0)}$, we obtain initial estimates of ${\bbeta_j}$ for the remaining items via standard sparse regression.
\end{remark}

 \subsection{Debiased Estimator}

% \textcolor{red}{Probably write down a hypothesis problem and write one sentence explaining why it is challenging (l1 penalty introduce bias, and cite a bunch of debiasing papers?) before you talk about debiasing?}

In this section, we focus on statistical inference for the covariate effects. Specifically, we let $\beta_{jk}$ denote the effect of the $k$-th observed covariate on the response to the $j$-th item and consider testing the null hypothesis $H_0: \beta_{jk}^* = 0$. A direct inferential procedure based on the regularized estimators from the alternating algorithm is generally not valid. A similar issue arises in high-dimensional regression without latent variables, where the $L_1$
 penalty introduces a non-negligible bias, and debiasing techniques are developed to address this challenge~\citep{zhang2014, van2014, javanmard2014confidence, ning2017}. However, in the presence of latent variables,  the problem is further complicated by the additional uncertainty arising from latent variable estimation, and whether traditional debiasing procedures remain applicable is underexplored.

% not only because of the non-negligible bias introduced by $L_1$ penalty, but also due to the presence of latent variable. With latent variables, it remains an open problem the applicable of traditional debiasing procedure due to the uncertainty of latent variable estimation 
% see, for example~\citet{zhang2014, van2014, javanmard2014confidence, ning2017}. 

% (1) due to the 

% (2) due to presence of latent variable estimation
% traditional methods without latent variables ...

% With latent variables, it remains an open problem the applicable of traditional debiasing procedure due to the uncertainty of latent variable estimation 
% see, for example~\citet{zhang2014, van2014, javanmard2014confidence, ning2017}. 

To address this challenge, we build on the estimators obtained from the alternating updates in~\eqref{eq:AM1}--\eqref{eq:AM2} and construct a debiased estimator $\tilde{\beta}_{jk}$, as summarized in Algorithm~\ref{alg:debiased-inference}. 
In particular, we take
the $t$-step estimators
 $\hat{\bbeta}_j^{(t)}$'s, $\hat{\beta}_{j0}^{(t)}$'s, $\hat\bgamma_j^{(t)}$'s and $\hat{\bU}_i^{(t)}$'s  of alternating algorithm 
 as the inputs to the debiasing procedure.

%In particular, we let $\hat{\bbeta}_j^{(t)}$'s, $\hat{\beta}_{j0}^{(t)}$'s, $\hat\bgamma_j^{(t)}$'s and $\hat{\bU}_i^{(t)}$'s be the $t$-step estimators given by the alternating algorithm in~\eqref{eq:AM1}--\eqref{eq:AM2} as the input. 

% \textcolor{red}{If you write down the hypothesis problem you do not need to restate this sentence here. Feel like it should be more clear if you say it directly what you are interested before introducing the procedure, since the procedure is based on debiasing $\beta_{jk}$ anyway.}

%We construct the debiased estimator using $\hat\bbeta_j$. 

\begin{algorithm}[H]
\caption{Debiasing procedure}
\label{alg:debiased-inference}

\KwIn{
$\{(y_{ij},\bX_{i}, \hat{\bU}_i^{(t)})\}_{i=1}^n$; $\hat\bvarphi_j =(\hat{\beta}_{j0}^{(t)}, (\hat{\bbeta}_j^{(t)})^{\T}, (\hat{\bgamma}_j^{(t)})^{\T})^{\T}$; covariate index $k$.

}

%\BlankLine
\textbf{Step 1}: Define the empirical loss $\hat\cL_j(\bvarphi_j) = -n^{-1}\sum_{i=1}^n l_{ij}(\beta_{j0} + \bbeta_j^{\T}\bX_i + \bgamma_j^{\T}\hat{\bU}_i^{(t)})$ and write $\bxi_j = (\beta_{j0}, \bbeta_{j,-k}^\T, \bgamma_j^\T)^\T$.
We compute $\hat{\wb}_j$ by
\begin{align}
    \hat{\wb}_j
=\operatorname*{argmin}_{\wb_j}\;
\frac{1}{2n}\!\left\{
\wb_j^{\T}\,\partial_{\bxi_j\bxi_j}\hat{\cL}_{j}(\hat\bvarphi_j)\,\wb_j
-2\wb_j^{\T}\,\partial_{\bxi_j\beta_{jk}}\hat{\cL}_{j}(\hat\bvarphi_j)
\right\}
+\lambda'\|\wb_j\|_1. \label{eq:estimate w}
\end{align}

\textbf{Step 2}:
Write $\hat\bZ_i = (1, \bX_i^{\T}, (\hat\bU_i^{(t)})^{\T})^{\T}$ and $\hat{\bZ}_{i, -k} = (1, \bX_{i,-k}^{\T}, ({\bU}_i^{(t)})^{\T})^{\T}$. Based on $\hat{\wb}_j$, compute
\begin{align*}
   \hat  S_j (\hat\beta_{jk}, \hat\bxi_j) &= \frac{1}{n}\sum_{i=1}^n l_{ij}^{\prime} (\hat{\bvarphi}_j^{\T} \hat{\bZ}_i) (X_{ik}  - \hat\wb_j^{\T}\hat{\bZ}_{i,-k}); \\
      \hat{F}_{j, k | \xi}  & = \frac{1}{n}\sum_{i=1}^n l_{ij}^{\prime\prime} (\hat{\bvarphi}_j^{\T} \hat{\bZ}_i)X_{ik} (X_{ik}  - \hat\wb_j^{\T}\hat{\bZ}_{i,-k}).
\end{align*}

\textbf{Step 3}: Set $\tilde{\beta}_{jk}
=\hat{\beta}_{jk}-\big(\hat{F}_{j, k | \xi} \big)^{-1}
\hat  S_j (\hat\beta_{jk}, \hat\bxi_j)$.

\KwOut{Debiased estimator $\tilde{\beta}_{jk}$.}

\end{algorithm}

%We assume that the effect of the $l$-th covariate to response to the $j$-th item is our parameter of interest, denoted as $\beta_{jl}$. We construct debiased estimator $\tilde{\beta}_{jl}$ for the parameter of interest and the debiasing procedure is presented in Algorithm~\ref{alg:debiased-inference}. In Section~\ref{sec:theory}, we will show that this estimator is asymptotically normal (see Theorem~\ref{thm:ind asymp normal}), which enables valid confidence intervals and hypothesis tests for parameter of interest $\beta_{jl}^*$.

%{\chengyu Remove:}We consider the effect of the $k$-th observed covariate on the response to the $j$-th item, denoted by $\beta_{jk}$, as the parameter of interest. The effects of other covariates effects are denoted as $\bbeta_{j, -k}$. We construct a debiased estimator $\tilde{\beta}_{jk}$ for $\beta_{jk}$ via the debiasing procedure summarized in Algorithm~\ref{alg:debiased-inference}. In Section~\ref{sec:theory}, we will show that $\tilde{\beta}_{jk}$ is asymptotically normal (see Theorem~\ref{thm:ind asymp normal}), which enables valid confidence intervals and hypothesis tests for $\beta_{jk}^*$.
Our approach is motivated by the debiasing methods for
high-dimensional inference without latent variables~\citep{ning2017}. Generally,
we remove the bias by augmenting the initial estimator with a correction term based on an approximate inverse of the derivative of the local score, where the estimated latent variables are used as surrogates for the unobserved latent variables.
%(Describe the algorithm from a high level. Elaborate with a few sentences of how to solve $\argmin_{\wb_j}$, as for Equations~\eqref{eq:AM1} and~\eqref{eq:AM2})
In Section~\ref{sec:theory}, we show that
$\tilde{\beta}_{jk}$ is asymptotically normal in Theorem~\ref{thm:ind asymp normal},
enabling valid confidence intervals for the true covariate effect.

\section{Theoretical Results}
\label{sec:theory}
In this section, we establish consistency of the proposed estimators and derive the asymptotic distribution of the debiased estimator $\tilde{\beta}_{jk}$ for the covariate effect of interest. We first state the assumptions needed for the theoretical analysis. For notational convenience, we use the superscript $^*$ to denote true parameter values. %let $(\bbeta^*_0,\Bb^*,\bGamma^*,\Ub^*)$ denote the true model parameters.

\begin{assumption}
	\label{assumption: psd covariance}
There exist constants $M >0$ and $\kappa  >0$ such that:
(i) The matrix $\bSigma_u^* = \lim_{n\rightarrow \infty} $ $ n^{-1} \Ub^{*\intercal} \Ub^* $ exists and is positive definite, with $\lambda_{\min} (\bSigma_u^*) \geq \kappa$. In addition, $\|\bU_i^*\|_{2} \le M$ for all $i \in [n]$;
(ii) The matrix $\bSigma_\gamma^* = \lim_{q \rightarrow \infty} q^{-1}\bGamma^{*\intercal} \bGamma^* $ exists and is positive definite, with $\lambda_{\min} (\bSigma_{\gamma}^*) \geq \kappa$. In addition, $\|\bgamma_j^*\|_{2} \le M$ for all $j \in [q]$;
(iii)  
$\EE(\bX_i) = \bm{0}$, $\|\bX_i\|_{\infty} \le M$, and $\max_{i\in[n],j\in[q]} $ $|(\bbeta_j^*)^{\T} \bX_i |\le M$;
(iv) Define $\bSigma_x = \EE(\bX_i \bX_i^{\T})$ and $\Fb_j^* = \EE\{l_{ij}^{\prime\prime}(\bvarphi_j^{*\T}\bZ_i^*)\, \bZ_i^* \bZ_i^{*\T} \}$\footnote{The expectation is evaluated under joint probability of independent and identical data $\{\by_i, \bX_i\}_{i=1}^n$.}, we have $\lambda_{\min} (\bSigma_x) \geq \kappa$,  and $\lambda_{\min}(\Fb_j^*) \geq \kappa$. 

	\end{assumption}
	
        Assumption~\ref{assumption: psd covariance} $(i)$--$(ii)$ correspond to Assumptions A-B in~\cite{bai2003inferential}. 
        The compact parameter space is also commonly assumed in nonlinear regression models~\citep{newey1994large}. 
Assumption~\ref{assumption: psd covariance} $(iii)$--$(iv)$ extend the standard regularity conditions for generalized linear models without latent variables to generalized latent variable models; see, for example, Assumption E.1 in \citet{ning2017} and related conditions in \cite{van2014}.

	\begin{assumption}
	\label{assumption:smoothness}
For $i \in [n]$ and $j \in [q]$, assume that $l_{ij}(\cdot)$ is three times differentiable. There exist $M > 0$ and large $\xi >2$ such that $\EE(|l_{ij}^{\prime}(w_{ij}^*) |^{\xi}) \le M$. $|l_{ij}^{\prime}(w_{ij}^*) |$ has mean zero and is sub-exponential with $\|l_{ij}^{\prime}(w_{ij}^*) \|_{\varphi_1} \le M$. There exists $b_U > b_L > 0 $ such that $b_L \le -l_{ij}^{\prime\prime}(w_{ij}) \le b_U$ within a compact space of $w_{ij}$ and $|l_{ij}^{\prime\prime}(w_{1})-l_{ij}^{\prime\prime}(w)| \le B |w_1 - w| |l_{ij}^{\prime\prime}(w)|$ for constant $B>0$, where $w \in [a_1 - \epsilon, a_2 + \epsilon]$ for $\epsilon > 0$ and sequence $w_1$ satisfies $|w_1 - w| = o(1)$. Finally, $|l_{ij}^{(3)}(w_{ij})| \le b_U$  within a compact space of $w_{ij}$. 
 	
			\end{assumption}

Assumption~\ref{assumption:smoothness} imposes standard regularity conditions on the log-likelihood function $l_{ij}(w_{ij})$. The boundedness assumption is required to guarantee the convexity of the objective functions in~\eqref{eq:AM1}--\eqref{eq:AM2}.
Under the commonly used generalized linear model setting, such a condition is mild. 
For example, in the logistic latent variable model, $l_{ij}^{\prime}(w_{ij}) = Y_{ij} - \exp(w_{ij})/\{1+\exp(w_{ij})\}$ and it follows $|l_{ij}^{\prime}(w_{ij})|\le 1$. We can similarly verify that $|l_{ij}^{\prime\prime}(w_{ij})|$ and $|l_{ij}^{(3)}(w_{ij})|$ are also uniformly bounded, and therefore Assumption~\ref{assumption:smoothness} holds naturally. For linear, probit, and Poisson factor models, Assumption~\ref{assumption:smoothness} can also be easily verified.

In alternating algorithm, the estimation errors of $\hat{\bbeta}_j^{(1)}$, $\hat{\beta}_{j0}^{(1)}$, and $\hat{\bgamma}_j^{(1)}$ depend on the estimation error of the initial estimator $\hat{\Ub}^{(0)}$, denoted by $\varepsilon_{nq}^{(0)}$. We therefore impose the following condition on the initialization. Let $s_{p,j} = \text{card}\{l:\beta_{jl}^* \neq 0\}$ be the support size of the covariate effects for item $j$, that is, the number of covariates with nonzero effects for item $j$. Define the overall sparsity level as $s_p = \max_{j\in[q]}s_{p,j}$.

\begin{assumption}
\label{assumption:intialization}
There exists an invertible matrix $\Gb \in \RR^{K \times K}$ such that the initial estimator $\hat{\Ub}^{(0)}$ satisfies $n^{-1/2} \| \hat{\Ub}^{(0)}\Gb^{\T} - \Ub^*\|_F
= O_p ( \varepsilon_{nq}^{(0)} )$
with $\varepsilon_{nq}^{(0)}$ satisfying $s_p\, \varepsilon_{nq}^{(0)} = o(1)$.
\end{assumption}

Assumption~\ref{assumption:intialization} requires the initial estimator of the latent variable to be average consistent up to an invertible transformation. This condition is mild and is satisfied by various constructions of the initial estimator \(\hat{\Ub}^{(0)}\).
For instance, the illustrative initialization example in~\eqref{eq:initial estimation} satisfies this assumption, as shown in the following proposition.
\begin{proposition}
\label{prop:initialization}
   Under Assumptions~\ref{assumption: psd covariance}--\ref{assumption:smoothness}, the initial estimators $\hat{\Ub}^{(0)}$ and $\hat{\bGamma}^{(0)}$ in~\eqref{eq:initial estimation} satisfy
$n^{-1/2}\| \hat{\Ub}^{(0)}\Gb^{\T} - \Ub^* \|_F = O_p ( \sqrt{s_q s_p/(n\wedge q)})$ and $q^{-1/2} \| \hat{\bGamma}^{(0)}\Gb^{-1} - \bGamma^* \|_F = O_p ( \sqrt{s_q s_p/(n\wedge q)})$, for some invertible matrix $\Gb \in \RR^{K \times K}$, where
$s_q = \text{card}\{j:\beta_{jl}^* \neq 0\text{ for some }l \in [p]\}$ and satisfies $s_q s_p^3 = o(n \wedge q)$.

%[check the rate includes $\sqrt{s_q s_p/(n\wedge q)}$]

\end{proposition}
Here $s_q$ is the number of items that exhibit a non-zero effect for at least one covariate. The average convergence rates of $ \hat{\Ub}^{(0)}\Gb^{\T} $ and $\hat{\bGamma}^{(0)}\Gb^{-1}$ are of the same order as those for the estimated latent variables and loadings from generalized factor models~\citep{wang2022maximum}. 
Moreover, for the initialization in~\eqref{eq:initial estimation}, this proposition shows that $\varepsilon_{nq}^{(0)} \asymp \sqrt{{s_q s_p}/{(n\wedge q)}}$ and Assumption~\ref{assumption:intialization} holds under 
the condition $s_q s_p^3 = o(n \wedge q)$. 
This sparsity requirement is mild. In particular, when $s_p$ is of constant order, a sensible setting in our motivating applications in educational and psychological assessments, the condition reduces to $s_q = o(n \wedge q)$, so the number of items with nonzero effects may increase with $n$ and $q$ at a sublinear rate. Note that our setting differs from standard sparsity setting in high-dimensional regression problems; see e.g., Chapter 7 in~\cite{wainwright2019high}. % such as requiring the sparsity level to be of order $o(\sqrt{n} / \log p)$.

Given these assumptions, we next establish the estimation errors of the estimators from the alternating algorithm in Theorem~\ref{thm: hd convergence rates}.

\begin{theorem}[Estimation Consistency]
\label{thm: hd convergence rates}
	Under Assumptions~\ref{assumption: psd covariance}--\ref{assumption:intialization}, at iteration $t=1$, suppose the tuning parameter in~\eqref{eq:AM1} satisfies $\lambda^{(1)} \asymp \sqrt{n^{-1}\log p} + \varepsilon_{nq}^{(0)}$. 

\noindent {\bf\small Results for~\eqref{eq:AM1}}: There exists some invertible matrix $\Gb$ such that, for any $j \in [q]$, we have: $\|\Gb^{-\T} \hat{\bgamma}_j^{(1)} - \bgamma_j^*\|_1 = O_p \{s_p^{1/2}(\sqrt{n^{-1} {\log p}} + \varepsilon_{nq}^{(0)})\}$, $|\hat{\beta}_{j0}^{(1)} - \beta_{j0}^*| = O_p \{s_p(\sqrt{n^{-1} {\log p}} +\varepsilon_{nq}^{(0)})\}$,   
	\begin{align*}
	  \| \hat{\bbeta}_j^{(1)} - \bbeta_j^*\|_1 &= O_p \Big\{s_p\big(\sqrt{\frac{\log p}{n}} + \varepsilon_{nq}^{(0)}\big)\Big\}, \quad 
 \| \hat{\bbeta}_j^{(1)} - \bbeta_j^*\|_2  = O_p \Big\{s_p^{1/2}\big(\sqrt{\frac{\log p}{n}} + \varepsilon_{nq}^{(0)}\big)\Big\}.
	\end{align*}
   In addition, we have $q^{-1/2}\|\hat{\bGamma}^{(1)}\Gb^{-1} - \bGamma^*\|_F = O_p $ $ \{s_p^{1/2}(\sqrt{n^{-1} {\log p}} + \varepsilon_{nq}^{(0)})\}$ and $q^{-1/2}\|\hat{\Bb}^{(1)} - \Bb^*\|_F = O_p $ $ \{s_p^{1/2}(\sqrt{n^{-1} {\log p}} + \varepsilon_{nq}^{(0)})\}$.

\noindent {\bf\small Results for~\eqref{eq:AM2}}: 
Given the above results for~\eqref{eq:AM1}, we further have the following results:
     \begin{align*}
\max_{i\in[n]} \| \Gb \hat{\bU}_i^{(1)} - \bU_i^*\|_{\infty} &= O_p \Big\{ \sqrt{\frac{\log n}{q}} + s_p^{1/2} \Big(\sqrt{\frac{\log p}{n}} + \varepsilon_{nq}^{(0)}\Big)\Big\}.
\end{align*}
    % In addition, we have 
    %  $n^{-1/2} \|\hat{\Ub}^{(1)}\Gb^{\T} - \Ub^*\|_{F} = O_p \{\sqrt{q^{-1}\log n}+s_p^{1/2}(\sqrt{n^{-1} {\log p}} + \varepsilon_{nq}^{(0)})\}$. 
\end{theorem}

%[separate the discussion points into remarks to avoid long texts]

Theorem~\ref{thm: hd convergence rates} shows that, under standard regularity assumptions, consistency of the initial estimators for latent variables implies the consistency of estimators for covariate effects and loadings.
More specifically, the first component in the individual convergence rates of $\hat{\bbeta}_j^{(1)}$ and of $\Gb^{-\T}\hat{\bgamma}_j^{(1)}$ in $L_1$ norm is of order $s_p\sqrt{(\log p)/n}$, matching the optimal convergence rate for regression coefficient estimators in high-dimensional regression models without latent variables~\citep{zhang2014, van2014, ning2017}. 
The second component is $\varepsilon_{nq}^{(0)}$, which quantifies the effect of estimating the latent variables and satisfies the scaling assumption that $ s_p\varepsilon_{nq}^{(0)} = o(1)$.

Once the item-specific
parameters are estimated at the rates given in
Theorem~\ref{thm: hd convergence rates}, the updated latent variable estimator from~\eqref{eq:AM2} also achieves consistency uniformly over subjects.
 Specifically, Theorem~\ref{thm: hd convergence rates} provides a bound on the maximum entrywise deviation between $\Gb \hat{\bU}_i^{(1)}$ and $\bU_i^*$ uniformly over all subjects.
 The first term $\sqrt{(\log n)/q}$ arises from simultaneously controlling this deviation over all subjects, while the second term is inherited from the estimation error of the item-specific
parameters obtained from~\eqref{eq:AM1}.
 This uniform consistency result plays a key role in establishing the asymptotic normality of the debiased estimator $\tilde{\beta}_{jk}$.

 %: the estimation error at each step is controlled by the standard high-dimensional term and the quality of the latent variable estimate carried over from the previous step.

% [Discuss the $s_p$ representation in the high-dimensional multivariate, $q*s_p$ out of $q*p$, this is not for only one $j$, if view it from overall picture, the number of nonzero coefficients is large.. even for $s_p \sim \sqrt{n}$, $q \sim n \sim p$, $n^{3/2}$ out of $n^2$ 

% Different from high-dimensional regression problem as the asymptotic regime is different, we are actually more dense, in the multivariate regression case, (blessing from $q$ dimensionality), than the high-dim univariate regression.
% ]

\begin{remark}
The required scaling condition on the sparsity level $s_p$ is mild. Although each item-specific \(\bbeta_j^*\) is allowed to have at most \(s_p\) nonzero entries, the
full coefficient matrix \(\Bb^*\in\RR^{q\times p}\) as a whole may contain as many as \(q \times s_p\) nonzero
entries.
Therefore, in the multivariate response setting here, the overall signal structure in the coefficient matrix can be denser than that in conventional univariate high-dimensional regression, reflecting a blessing of response dimensionality.
To illustrate, consider  \(q\asymp n\) and
\(s_p\asymp n^{1/2}\), the total number of nonzero coefficients can be of order
\(n^{3/2}\) out of \(n^2\) entries of $\Bb^*$. In this sense, while sparsity is imposed at the row level, the global coefficient matrix is allowed to be considerably dense. 
\end{remark}

Furthermore, the results can be generalized to every iteration $t \ge 1$.  If the latent variable estimator from iteration $t-1$ is consistent with error order $\varepsilon_{nq}^{(t-1)}$, then the estimators $\hat{\bbeta}_j^{(t)}$, $\Gb^{-\T}\hat{\bgamma}_j^{(t)}$, and $\hat{\beta}_{j0}^{(t)}$ are also consistent. Given these results, we further have uniform and average consistency results for latent variables $\hat{\bU}_i^{(t)}$.
This yields a recursive characterization of the estimation error for algorithm updates across iterations.

\begin{corollary}
\label{coro:estimation t ge 1}
Under the conditions of Theorem~\ref{thm: hd convergence rates}, for any iteration $t > 1$, define $\varepsilon_{nq}^{(t-1)}$ and $\lambda^{(t)}$ analogously to $\varepsilon_{nq}^{(0)}$ and $\lambda^{(1)}$, where $\varepsilon_{nq}^{(t-1)}$ is a sequence such that
$n^{-1/2} \|\hat{\Ub}^{(t-1)}\Gb^{\T} - \Ub^*\|_{F} = O_p(\varepsilon_{nq}^{(t-1)})$ and $\lambda^{(t)} \asymp $ 
$\sqrt{n^{-1}\log p} + \varepsilon_{nq}^{(t-1)}$. Then the convergence results for the updates of~\eqref{eq:AM1} given in Theorem~\ref{thm: hd convergence rates} hold with $\varepsilon_{nq}^{(0)}$ replaced by $\varepsilon_{nq}^{(t-1)}$.
Similarly, given the corresponding results for the estimators from~\eqref{eq:AM1}, the convergence results for the update in~\eqref{eq:AM2} also hold with $\varepsilon_{nq}^{(0)}$ replaced by $\varepsilon_{nq}^{(t-1)}$.

\end{corollary}

In terms of practical implementation, we consider approximate algorithms, such as gradient-type methods, for~\eqref{eq:AM1} and~\eqref{eq:AM2}. We quantify the number of inner iterations needed for the resulting estimator to achieve the same convergence rate as the corresponding exact minimizer. 
Specifically, let $(\bar{\beta}_{j0}^{(t, M_1)}, \bar{\bbeta}_j^{(t, M_1)}, \bar{\bgamma}_j^{(t, M_1)})$ be the approximate minimizer of~\eqref{eq:AM1} after $M_1$ inner steps, and let $\bar{\bU}_i^{(t, M_2)}$ be the approximate minimizer of~\eqref{eq:AM2} after $M_2$ inner steps.
We next present Proposition~\ref{prop:AM1 num steps} to specify the requirements on the number of inner iterations for approximate algorithms.

\begin{proposition}
\label{prop:AM1 num steps}
  Suppose Assumptions~\ref{assumption: psd covariance}--\ref{assumption:intialization} hold.\\ (i) For each $j \in [q]$, let the proximal-gradient step size satisfy $\eta_{\phi} \leq 1/ L_j$, where $L_j$ is the Lipschitz constant of gradient of the loss in~\eqref{eq:AM1}. After $M_1$ number of inner iterations with $ M_1 \ge C_1\log n$,  the resulting estimator $(\bar{\beta}_{j0}^{(t, M_1)}, \bar{\bbeta}_j^{(t, M_1)}, \bar{\bgamma}_j^{(t, M_1)})$ in Algorithm~1 of Supplementary Materials satisfies the same convergence rate as the exact minimizer $(\hat{\beta}_{j0}^{(t)}, \hat{\bbeta}_j^{(t)}, \hat{\bgamma}_j^{(t)})$ in Theorem~\ref{thm: hd convergence rates}.\\
(ii) For $i \in [n]$, let the gradient step size satisfy $\eta_{u} \le 1/D_i$, where $D_i$ is the Lipschitz constant of gradient of the loss in~\eqref{eq:AM2}. After $M_2$ number of inner iterations with $M_2 \ge C_2 (\log n +\log q )$, the resulting estimator $\bar{\bU}_i^{(t, M_2)}$ in Algorithm~2 in Supplementary Materials satisfies the same convergence rate as the exact minimizer $\hat{\bU}_i^{(t)}$ in Theorem~\ref{thm: hd convergence rates}.
\end{proposition}

Proposition~\ref{prop:AM1 num steps} provides a computational guarantee for implementing the alternating algorithm using gradient-type approximate algorithms. With $M_1$ and $M_2$ inner iterations satisfying $ M_1 \ge C_1\log n$ and $M_2 \ge C_2 (\log n +\log q )$ for sufficiently large constants $C_1>0$ and $C_2>0$, the optimization errors of these approximate estimators decay fast and become asymptotically dominated by the statistical error. As a result, the estimators produced by these approximate algorithms attain the same convergence rates as those based on exact minimizers of~\eqref{eq:AM1} and~\eqref{eq:AM2}, as given in Theorem~\ref{thm: hd convergence rates}.

We next establish the asymptotic normality of the debiased estimator given in Algorithm~\ref{alg:debiased-inference}, thereby providing valid inference results. 
Building on the debiasing literature on high-dimensional inference~\citep{ning2017}, let
$\wb_j^*$ denote the population decorrelation vector for inference on $\beta_{jk}$; see Section A of the Supplementary Materials for the formal definition. As shown in Proposition A2 of Supplementary Materials, the estimator $\hat{\wb}_j$ in~\eqref{eq:estimate w} of Algorithm~\ref{alg:debiased-inference} is a consistent estimator of $\wb_j^*$.

\begin{theorem}[Asymptotic Normality]
\label{thm:ind asymp normal}
%Suppose the debiased estimator is constructed using the 
Consider the debiased estimator in Algorithm~\ref{alg:debiased-inference} with $t=1$. 
%input to be the first iteration estimator $t=1$.
% that is, $\hat{\bbeta}_j :=\hat{\bbeta}_j^{(1)}, \hat{\beta}_{j0}:= \hat{\beta}_{j0}^{(1)}, \hat{\bgamma}_j := \hat{\bgamma}_j^{(1)}$ and $\hat{\bU}_i := \hat{\bU}_i^{(1)}$. 
% estimators obtained after the first iteration of the alternating algorithm, that is, $\hat{\bbeta}_j:=\hat{\bbeta}_j^{(1)}, \hat{\beta}_{j0}:= \hat{\beta}_{j0}^{(1)}, \hat{\bgamma}_j:= \hat{\bgamma}_j^{(1)}$ and $\hat{\bU}_i:= \hat{\bU}_i^{(1)}$. 
Under Assumptions~\ref{assumption: psd covariance}--\ref{assumption:intialization}, assume there exist a constant $M >0$ such that $\|\wb_{j}^*\|_{\infty} \leq M$, and $\max_{i\in[n],j\in[q]}|(\wb_{j}^*)^{\T} \bZ_{i,-k}^*| \le M$.
Suppose the tuning parameters in~\eqref{eq:AM1} and in~\eqref{eq:estimate w} satisfy $\lambda^{(1)} \asymp \sqrt{n^{-1}\log p} + \varepsilon_{nq}^{(0)}$ and $ \lambda^{\prime}\asymp q^{-1/2} (\log n)^{1/2} +  s_p^{1/2}( \sqrt{n^{-1}\log p} + \varepsilon_{nq}^{(0)})$. Suppose \(n,q,p \to \infty\), let $\zeta_{nqp}^{(0)} = \sqrt{n^{-1}\log p} +  \varepsilon_{nq}^{(0)}$ and $s_j^* = \text{card} (\{l:w_{jl}^* \neq 0\})$. If $n^{1/2}\zeta_{nqp}^{(0)} (s_j \vee s_p)(\sqrt{q^{-1} \log n} + s_p^{1/2}\zeta_{nqp}^{(0)}) = o_p(1)$, then
\begin{equation}
	 n^{1/2}  ({F}_{j,k | \xi }^*)^{1/2}( \tilde{\beta}_{jk} - \beta_{jk}^* ) \xrightarrow{d} \mathcal{N} (0, 1). \nonumber
\end{equation}

Moreover, for any $t>1$,
% debiased estimator $\tilde{\beta}_{jk}$ constructed by final estimators after iterations $t \ge 1$, 
the same conclusion holds with $\varepsilon_{nq}^{(0)}$ and $\lambda^{(1)}$ replaced by
$\varepsilon_{nq}^{(t-1)}$ and $\lambda^{(t)}$ respectively, as defined in Corollary~\ref{coro:estimation t ge 1}, and with $\lambda^{\prime} \asymp  q^{-1/2} (\log n)^{1/2} +  s_p^{1/2} (\sqrt{n^{-1}\log p} + \varepsilon_{nq}^{(t-1)})$.

\end{theorem}

The results in Theorem~\ref{thm:ind asymp normal} are analogous to existing inference results for high-dimensional regression models without latent variables~\citep{zhang2014, van2014, ning2017}. 
The theorem requires the scaling condition $n^{1/2}\zeta_{nqp}^{(0)} (s_j \vee s_p)(\sqrt{q^{-1} \log n} + s_p^{1/2}\zeta_{nqp}^{(0)}) = o_p(1)$, which depends on $\zeta_{nqp}^{(0)}$ and hence on $\varepsilon_{nq}^{(0)}$, the error induced by estimating the latent variables in the alternating algorithm. In particular, this condition implies $\varepsilon_{nq}^{(0)} = o(n^{-1/4}(s_j s_p^{1/2} \vee s_p^{3/2} )^{-1/2} ) $ and $\varepsilon_{nq}^{(0)} = o(q^{1/2} n^{-1/2}(\log n)^{-1/2} (s_j \vee s_p)^{-1} )$. 
This stronger requirement relative to Assumption~\ref{assumption:intialization} is natural, since establishing asymptotic normality typically requires tighter error control than establishing consistency alone.
It is verified that covariate-free initialization in~\eqref{eq:initial estimation} also satisfies this stronger scaling condition.

 Once asymptotic normality is established, we can perform hypothesis testing and construct confidence intervals for any covariate effect of interest $\beta_{jk}^*$.  
With the asymptotic normality results in Theorem~\ref{thm:ind asymp normal}, we can construct a
$100(1-\alpha)\%$ confidence interval for $\beta_{jk}^*$ as $[
\tilde{\beta}_{jk} - {z_{1-\alpha/2}}({n\,\hat F_{j,k\mid \xi} })^{-1/2},
\ \tilde{\beta}_{jk} + {z_{1-\alpha/2}}({n\,\hat F_{j,k\mid \xi}})^{-1/2}
]$ where $z_{1-\alpha/2}$ is the $(1-\alpha/2)$-quantile of the standard normal distribution.

\begin{remark}
This work focuses on inference for the covariate effects $\bbeta_j$. In many applications, however, uncertainty quantification for the latent variables $\bU_i$ and loading parameters $\bgamma_j$ is also of great interest to statisticians and practitioners, since these parameters often carry important scientific interpretations, for example, $\bU_i$ may represent the latent skills or abilities to be assessed in educational assessments and psychological measurements~\citep{skrondal2004generalized}.  Establishing inference results for $\bU_i$ and $\bgamma_j$ is challenging because their estimation depends on the estimation of $\bbeta_j$; consequently, valid uncertainty quantification for $\bU_i$ and $\bgamma_j$ must account for the uncertainty in $\bbeta_j$. The asymptotic distribution for $\bbeta_j$ in Theorem~\ref{thm:ind asymp normal} provides an important foundation for deriving asymptotic distributions for estimators of $\bU_i$ and $\bgamma_j$, extending the existing inference theory for nonlinear factor analysis to our high-dimensional generalized latent variable modeling~\citep{wang2022maximum, ouyang2025statistical}. We leave a full investigation of this direction for future work.
\end{remark}

\section{Simulation}
\label{sec:simulation}

%[add competing methods - direct glm (debiasing package)]

%[add MSE]

To illustrate the performance of the proposed method, we consider a generalized latent variable model with a logit link, i.e., $p_{ij} (y \mid w_{ij}) = \exp(w_{ij}y)/\{1+\exp(w_{ij})\}$, where $w_{ij} = \beta_{j0}^*+  (\bgamma_j^*)^{\T} \bU_i^* + (\bbeta_j^*)^{\T} \bX_i$. We apply our method to estimate each individual covariate effect $\beta_{jl}^*$ and test $H_0: \beta_{jl}^* = 0$ for $j \in [q]$ and $l \in [p]$.
%For notational convenience, let $\Bb = (\bbeta_1, \dots, \bbeta_q)_{q \times p}^{\intercal}$ to denote the covariate effect vectors across all items.
The data generation process is as follows. We let $n  \in \{100, 300, 500\} $, $p \in \{500, 1000\}$, $q \in \{100, 300\}$, and $K =3$. We set $\gamma_{jk}^* \sim \cN(0,1)$ and generate covariates $\bX_{i}$ and latent variables $\bU_{i}$ independently and identically from $(\bX_i \; \bU_i^*) \sim \cN (\bm{0}, \bSigma)$,
where $\bSigma_{ij} = \rho^{|i - j|}$ with $\rho \in \{0, 0.2,  0.8 \}$,  for all $i \in[n]$ and $j \in [q]$. %, $k, h \in [K]$ and $t \in [p]$. 
%We centralize $\Ub^*$ and ensures $(\Ub^*)^{\T} \bm{1}_n = \bm{0}$ holds.
For the coefficient matrix $\Bb^*$, we let $\beta_{jl}^* \sim \text{Unif}[a, a + 0.5]$ 
for $j = 1, ..., J$ and $l= 1, ..., s$, with $a \in \{0.3, 0.5\}$ controlling the signal strength and $J \in \{10, 30\}$ and $s \in \{10, 50\}$ indicating the density level (i.e., the number of nonzero entries) of coefficient $\Bb^*$. The remaining entries in $\Bb^*$ are set to zero and the intercept $\beta_{j0}^* =1$ for all $j \in [q]$.

In implementing our method, we choose the parameter $\lambda$ by 5-fold cross-validation. The cross-validation error is calculated by the mean squared error between the predicted response probability of $y_{ij}=1$ and the true $y_{ij}$ in each test set. 
As a baseline comparison, we fit $L_1$-penalized generalized linear models (GLM) that regress multivariate responses on the observed covariates only, without adjusting for latent variables. Specifically, the baseline method first fits a standard regularized GLM that ignores latent variables, and then constructs debiased estimators following the debiasing framework in the literature~\citep{ning2017}. We calculate the power of the test based on the non-zero entries in $\{\beta_{jl}: (j,l) \in [10] \times [10] \}$. The type I error of the test is based on the zero entries in $\{\beta_{jl}: (j,l) \in \{51, \dots,  60\} \times [10] \}$. The rejection proportions over the 100 simulations for our proposed method and the baseline method under $q = 300$ and $a=0.3$ are presented in Table~\ref{tab:type1_q300_a03}. 
In addition, we evaluate the accuracy in estimating the true $\Bb^*$ using the mean squared errors (MSEs). Due to space constraints, we defer the type I errors and power results for the remaining settings and the MSE results to Section~G of the Supplementary Materials. %We compare this approach with the proposed method, and the results are reported in Figures~A1--A4 of Supplementary Materials. 

 From Table~\ref{tab:type1_q300_a03}, the Type I errors of the proposed method are all close to the nominal level of 0.05 across all settings. In contrast, the baseline approach fails to control Type I error at 0.05, suggesting systematic estimation bias. Regarding power, our method yields higher as the sample size $n$ increases; the baseline method generally exhibits high statistical power. In addition, the proposed method achieves smaller MSEs than the baseline method (see Section~G.2 of the Supplementary Materials).
The poor type I error control from the baseline method is expected because it ignores the latent structure (equivalently, sets $\Ub =\zero $). The variation in the response that is truly driven by latent variables $\bU_i$'s is instead absorbed by observed covariates $\bX_i$'s, which inflate the estimated covariate effects. Consequently, some coefficients that are truly zero may be estimated as nonzero, leading to increased rejection proportions. Correspondingly, the empirical power of the baseline method is also typically very high, often close to one.

For the proposed method, comparing the type I error rates of $p=500$ with those of $p=1000$ in Table~\ref{tab:type1_q300_a03}, we observe that all type I errors are approximately 0.05, indicating that our method is robust to the increase in the covariate dimension.
%We also observe that the performance of our method is robust when the number of items $q$ increases from $q = 100$ (Table~\ref{tab:type1_q100_a03}) to $q=300$ (Table~\ref{tab:type1_q300_a03}). 
In addition, our method is stable when the number of items $J$ corresponding to non-zero covariate effects increases or the number of covariates with non-zero covariate effects $s$ increases. Lastly, regarding the correlation between $\Xb$ and $\Ub$, we observe that regardless of the dependence level $\rho$, the power increases to one as the sample size increases. However, for a fixed sample size, the highly correlated setting $(\rho = 0.8)$ generates lower power than the weakly correlated $(\rho = 0.2)$ and non-correlated settings $(\rho = 0)$, and also exhibits larger MSEs than the latter two cases (see details in Section~G.2 of the Supplementary Materials).

\begin{table}[!ht]
\centering
\footnotesize
\setlength{\tabcolsep}{1.6pt}
\renewcommand{\arraystretch}{0.6}
\begin{tabular}{cc| *{6}{S} @{\hspace{6pt}\vrule} *{6}{S}}
\toprule
\multicolumn{2}{c}{ } &  \multicolumn{6}{c|}{Proposed Method} &  \multicolumn{6}{c}{Baseline Method}  \\
 \cmidrule(lr){3-8} \cmidrule(lr){9-14}
\multicolumn{2}{c}{ } & \multicolumn{3}{c}{$p=500$} & \multicolumn{3}{c|}{$p=1000$} & \multicolumn{3}{c}{$p=500$} & \multicolumn{3}{c}{$p=1000$}  \\
\cmidrule(lr){1-2} \cmidrule(lr){3-5} \cmidrule(lr){6-8} \cmidrule(lr){9-11} \cmidrule(lr){12-14} 
%$J$ & $s$ & {$n=100$} & {$300$} & {$500$} & {$100$} & {$300$} & {$500$} & {$100$} & {$300$} & {$500$} & {$100$} & {$300$} & {$500$}  \\ \midrule
$J$ & $s$ &
\multicolumn{3}{c}{$n$} & \multicolumn{3}{c|}{$n$} &
\multicolumn{3}{c}{$n$} & \multicolumn{3}{c}{$n$} \\
\cmidrule(lr){1-2}\cmidrule(lr){3-5}\cmidrule(lr){6-8}\cmidrule(lr){9-11}\cmidrule(lr){12-14}
& &
{100} & {300} & {500} & {100} & {300} & {500} &
{100} & {300} & {500} & {100} & {300} & {500} \\
\multicolumn{14}{c}{\textit{Panel A: $\rho=0$\;\;\;}} \\
\addlinespace[1pt]
10 & 10 & 0.052 & 0.047 & 0.050 & 0.050 & 0.051 & 0.050 & 0.135 & 0.322 & 0.596 & 0.073 & 0.193 & 0.282 \\  
30 & 10 & 0.053 & 0.047 & 0.051 & 0.051 & 0.051 & 0.049 & 0.135 & 0.319 & 0.597 & 0.072 & 0.191 & 0.282 \\ 
10 & 50 & 0.053 & 0.045 & 0.052 & 0.050 & 0.051 & 0.050 & 0.134 & 0.321 & 0.596 & 0.073 & 0.194 & 0.286 \\
30 & 50 & 0.051 & 0.049 & 0.053 & 0.050 & 0.051 & 0.049 & 0.135 & 0.318 & 0.599 & 0.074 & 0.190 & 0.285 \\ 
\addlinespace[3pt]

\multicolumn{14}{c}{\textit{Panel B:  $\rho=0.2$}} \\
\addlinespace[1pt]
10 & 10 & 0.053 & 0.045 & 0.048 & 0.049 & 0.052 & 0.052 & 0.131 & 0.323 & 0.599 & 0.074 & 0.184 & 0.285 \\
30 & 10 & 0.052 & 0.045 & 0.050 & 0.049 & 0.051 & 0.050 & 0.133 & 0.325 & 0.590 & 0.075 & 0.186 & 0.284 \\ 
10 & 50 & 0.052 & 0.046 & 0.048 & 0.049 & 0.050 & 0.052 & 0.131 & 0.324 & 0.598 & 0.074 & 0.184 & 0.289 \\ 
30 & 50 & 0.051 & 0.048 & 0.051 & 0.049 & 0.049 & 0.051 & 0.131 & 0.322 & 0.599 & 0.076 & 0.180 & 0.282 \\ 
\addlinespace[3pt]

\multicolumn{14}{c}{\textit{Panel C:  $\rho=0.8$}} \\
\addlinespace[1pt]
10 & 10  & 0.051 & 0.050 & 0.049 & 0.049 & 0.051 & 0.049 & 0.128 & 0.324 & 0.612 & 0.074 & 0.178 & 0.277 \\  
30 & 10  & 0.051 & 0.048 & 0.048 & 0.051 & 0.049 & 0.048 & 0.131 & 0.327 & 0.606 & 0.073 & 0.179 & 0.275 \\ 
10 & 50  & 0.050 & 0.050 & 0.049 & 0.051 & 0.051 & 0.048 & 0.129 & 0.326 & 0.612 & 0.074 & 0.179 & 0.277 \\  
30 & 50  & 0.054 & 0.051 & 0.049 & 0.051 & 0.048 & 0.047 & 0.131 & 0.323 & 0.606 & 0.073 & 0.178 & 0.274 \\ 
\bottomrule
\\
\end{tabular}

\begin{tabular}{cc| *{6}{S} @{\hspace{6pt}\vrule} *{6}{S}}
\toprule
\multicolumn{2}{c}{ } &  \multicolumn{6}{c|}{Proposed Method} &  \multicolumn{6}{c}{Baseline Method}  \\
 \cmidrule(lr){3-8} \cmidrule(lr){9-14}
\multicolumn{2}{c}{ } & \multicolumn{3}{c}{$p=500$} & \multicolumn{3}{c|}{$p=1000$} & \multicolumn{3}{c}{$p=500$} & \multicolumn{3}{c}{$p=1000$}  \\
\cmidrule(lr){1-2} \cmidrule(lr){3-5} \cmidrule(lr){6-8} \cmidrule(lr){9-11} \cmidrule(lr){12-14} 
%$J$ & $s$ & {$n=100$} & {$300$} & {$500$} & {$100$} & {$300$} & {$500$} & {$100$} & {$300$} & {$500$} & {$100$} & {$300$} & {$500$}  \\ \midrule
$J$ & $s$ &
\multicolumn{3}{c}{$n$} & \multicolumn{3}{c|}{$n$} &
\multicolumn{3}{c}{$n$} & \multicolumn{3}{c}{$n$} \\
\cmidrule(lr){1-2} \cmidrule(lr){3-5}\cmidrule(lr){6-8}\cmidrule(lr){9-11}\cmidrule(lr){12-14}
& &
{100} & {300} & {500} & {100} & {300} & {500} &
{100} & {300} & {500} & {100} & {300} & {500} \\
\multicolumn{14}{c}{\textit{Panel A: $\rho=0$\;\;\;}} \\
\addlinespace[1pt]
10 & 10 & 0.745 & 0.960 & 0.994 & 0.738 & 0.960 & 0.995 & 0.946 & 0.948 & 0.934 & 0.957 & 0.978 & 0.974  \\  
30 & 10 & 0.740 & 0.939 & 0.993 & 0.713 & 0.965 & 0.993 & 0.939 & 0.945 & 0.934 & 0.957 & 0.977 & 0.973 \\ 
10 & 50 & 0.742 & 0.944 & 0.995 & 0.719 & 0.949 & 0.993 & 0.942 & 0.943 & 0.933 & 0.945 & 0.976 & 0.972 \\ 
30 & 50 & 0.736 & 0.941 & 0.993 & 0.713 & 0.962 & 0.993 & 0.936 & 0.944 & 0.936 & 0.947 & 0.979 & 0.975 \\ 
\addlinespace[3pt]

\multicolumn{14}{c}{\textit{Panel B:  $\rho=0.2$}} \\
\addlinespace[1pt]
10 & 10 & 0.741 & 0.962 & 0.994 & 0.740 & 0.960 & 0.997 & 0.944 & 0.945 & 0.937 & 0.960 & 0.976 & 0.975 \\
30 & 10 & 0.736 & 0.944 & 0.994 & 0.714 & 0.962 & 0.994 & 0.940 & 0.941 & 0.933 & 0.957 & 0.976 & 0.971 \\
10 & 50 & 0.739 & 0.946 & 0.994 & 0.720 & 0.943 & 0.993 & 0.941 & 0.937 & 0.933 & 0.947 & 0.975 & 0.973 \\
30 & 50 & 0.730 & 0.940 & 0.993 & 0.713 & 0.963 & 0.994 & 0.931 & 0.940 & 0.939 & 0.946 & 0.976 & 0.974 \\ 
\addlinespace[3pt]

\multicolumn{14}{c}{\textit{Panel C:  $\rho=0.8$}} \\
\addlinespace[1pt]
10 & 10 & 0.642 & 0.900 & 0.971 & 0.650 & 0.907 & 0.975 & 0.917 & 0.935 & 0.904 & 0.934 & 0.966 & 0.959 \\
30 & 10 & 0.645 & 0.861 & 0.968 & 0.627 & 0.911 & 0.970 & 0.916 & 0.928 & 0.903 & 0.929 & 0.963 & 0.958  \\ 
10 & 50  & 0.642 & 0.876 & 0.970 & 0.633 & 0.893 & 0.970 & 0.914 & 0.937 & 0.903 & 0.912 & 0.964 & 0.956  \\ 
30 & 50  & 0.642 & 0.861 & 0.968 & 0.639 & 0.911 & 0.969 & 0.903 & 0.929 & 0.902 & 0.911 & 0.969 & 0.956 \\ 
\bottomrule
\end{tabular}
\caption{Empirical Type I error rates (top panel) and empirical power (bottom panel) for the proposed method and baseline method for $q=300$ and $a=0.3$ over 100 simulations.}
\label{tab:type1_q300_a03}
\end{table}

\section{Data Application}
\label{sec:data application}
%\subsection{Application to PISA 2022}\newpage
We analyze data from the 2022 cycle of the Programme for International Student Assessment (PISA) to assess the effect of covariates, including country-of-origins and assessment languages, on item-level performances, after accounting for latent abilities. PISA is a large-scale international assessment that evaluates the cognitive skills of 15-year-old students in mathematics, science, and reading, across countries and economies worldwide~\citep{oecd2024}. In addition to item responses in the three domains, PISA 2022 collects comprehensive information on students' demographics (such as gender, race, and country of origin) as well as details about students' testing, learning, and living environments~\citep{oecd2024}.
%In 2022, more than 600,000 students from 81 countries participated in the PISA program. 

The PISA 2022 assessment includes $q=455$ items: 169 in mathematics, 103 in science, and 183 in reading.
The item responses are binary, where $y_{ij}=1$ indicates that student $i$ answered item $j$ correctly and $y_{ij}=0$ otherwise. We model the binary responses $y_{ij}$ using a logistic model with $p_{ij} (y \mid w_{ij}) = \exp(w_{ij}y)/\{1+\exp(w_{ij})\}$.
% To reduce the burden on students and test administration, PISA uses a matrix-sampling scheme, in which different groups of students answer different yet overlapping sets of items. Such a test setting creates a systematic missingness in the observed response matrix [refer to aoas paper].
 During data preprocessing, we observe that some items have no responses, and some students answer only a few items. To ensure a balanced pattern of observed responses, we remove items with fewer than two responses and students who answered fewer than 10 items.
We incorporate $p=151$ covariates, including gender, country of origin, grade level, parents’ education, and the language of the assessment. After data cleaning, the final sample consists of $n=7{,}591$ students. Following existing literature~\citep{schleicher2019pisa, oecd2024}, we set the latent dimension to $K=3$ to align with the three core domains of assessment: mathematics, reading, and science.

\begin{figure}
    \centering
    \includegraphics[width=\linewidth]{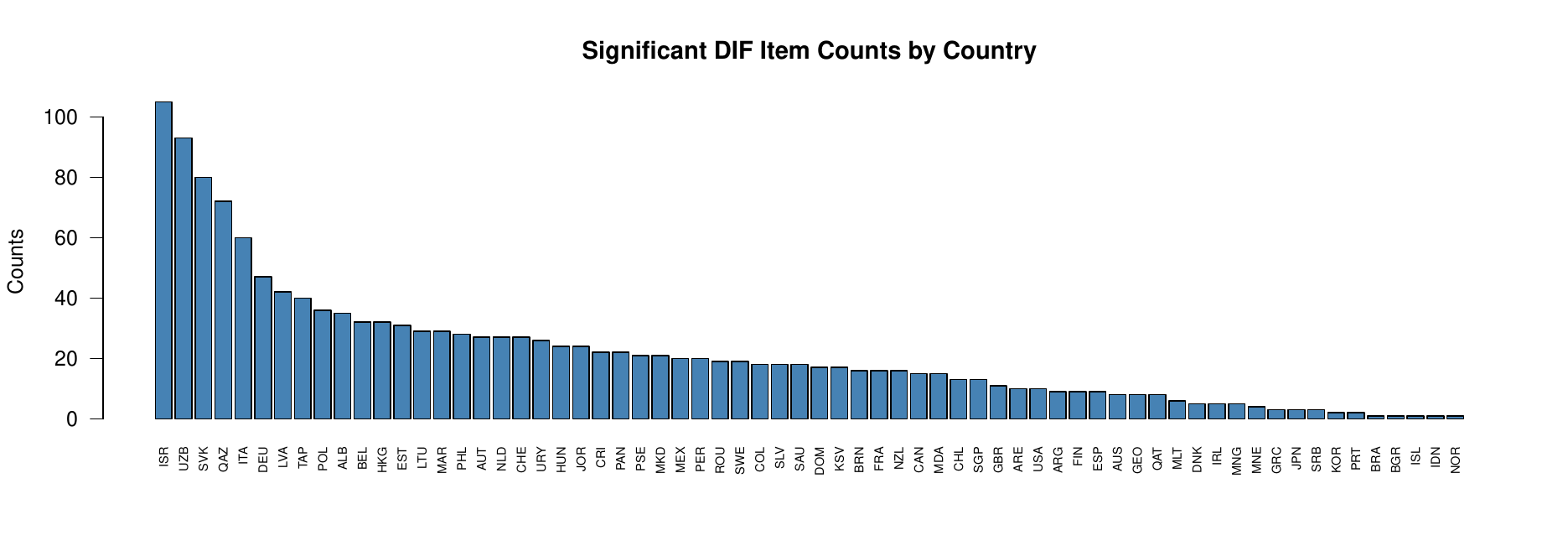}
    \caption{Number of Biased Items for each country in the PISA study.}
    \label{fig:countries}
\end{figure}

\begin{figure}[h!]
    \centering
    \includegraphics[width=0.5\linewidth]{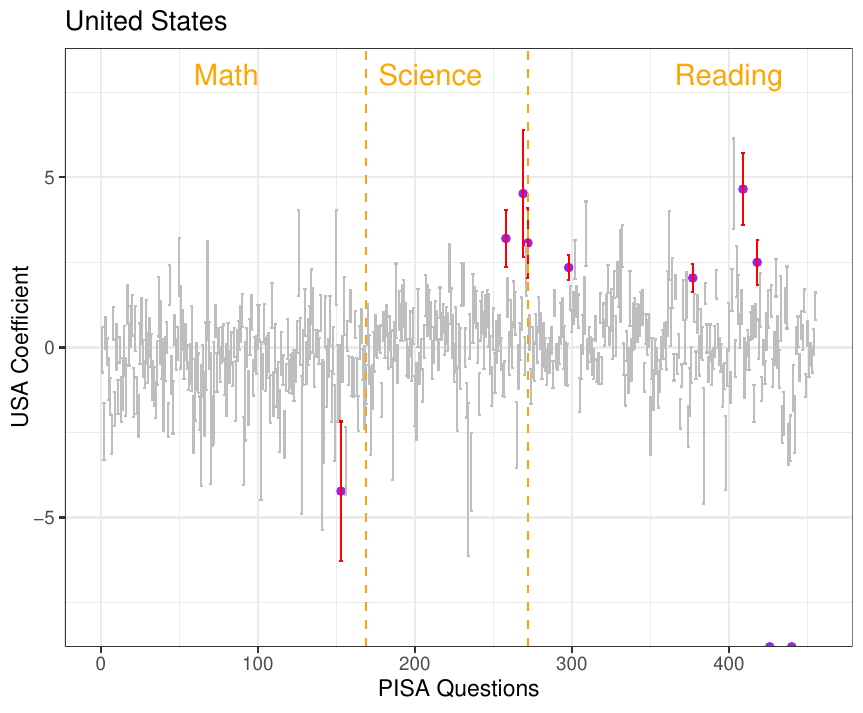}~\includegraphics[width=0.5\linewidth]{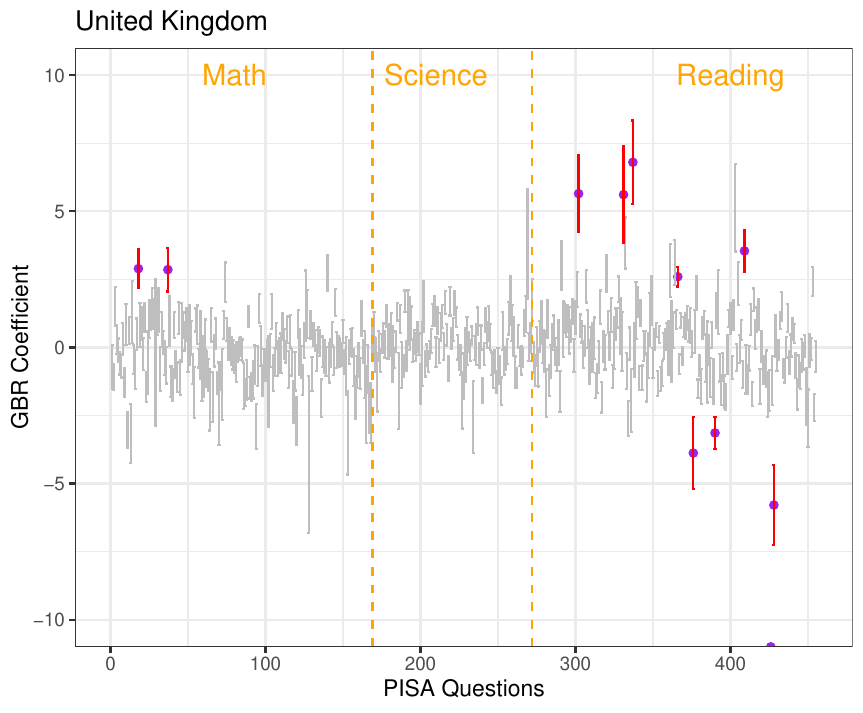}\\
    \includegraphics[width=0.5\linewidth]{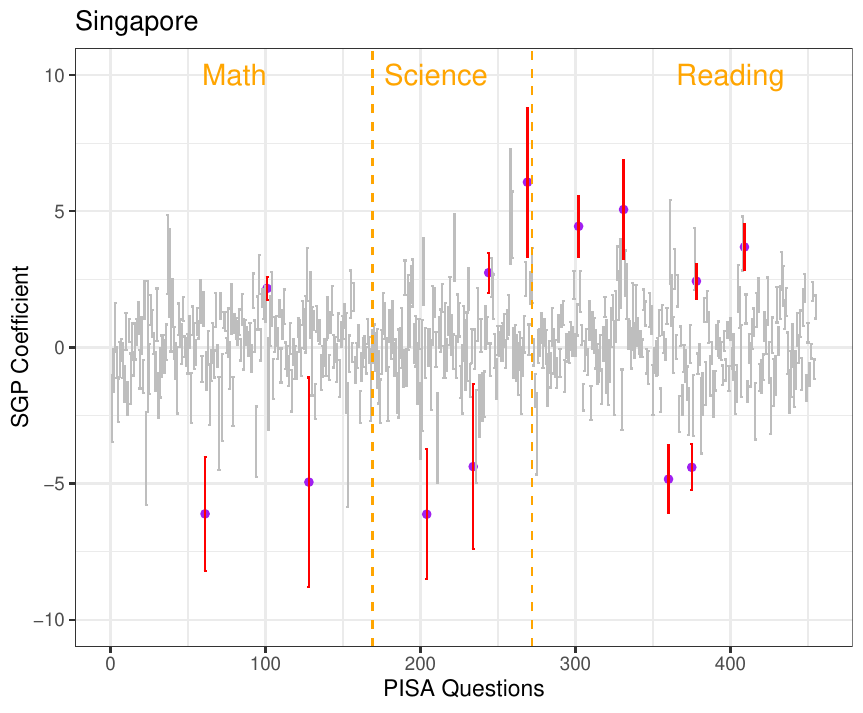}~
    \includegraphics[width=0.5\linewidth]{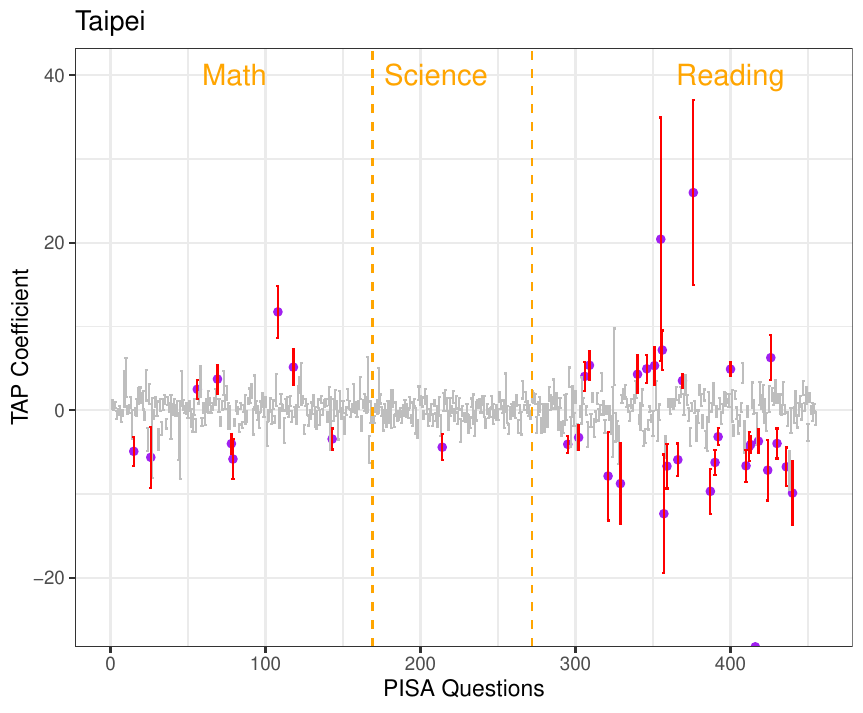}
    \caption{Confidence intervals for effects of selective country-of-origin indicators on each of PISA items. Red intervals correspond to confidence intervals for items with significant bias after Bonferroni correction.}
    \label{fig:country bias}
\end{figure}

\begin{figure}[h!]
    \centering
    \includegraphics[width=0.5\linewidth]{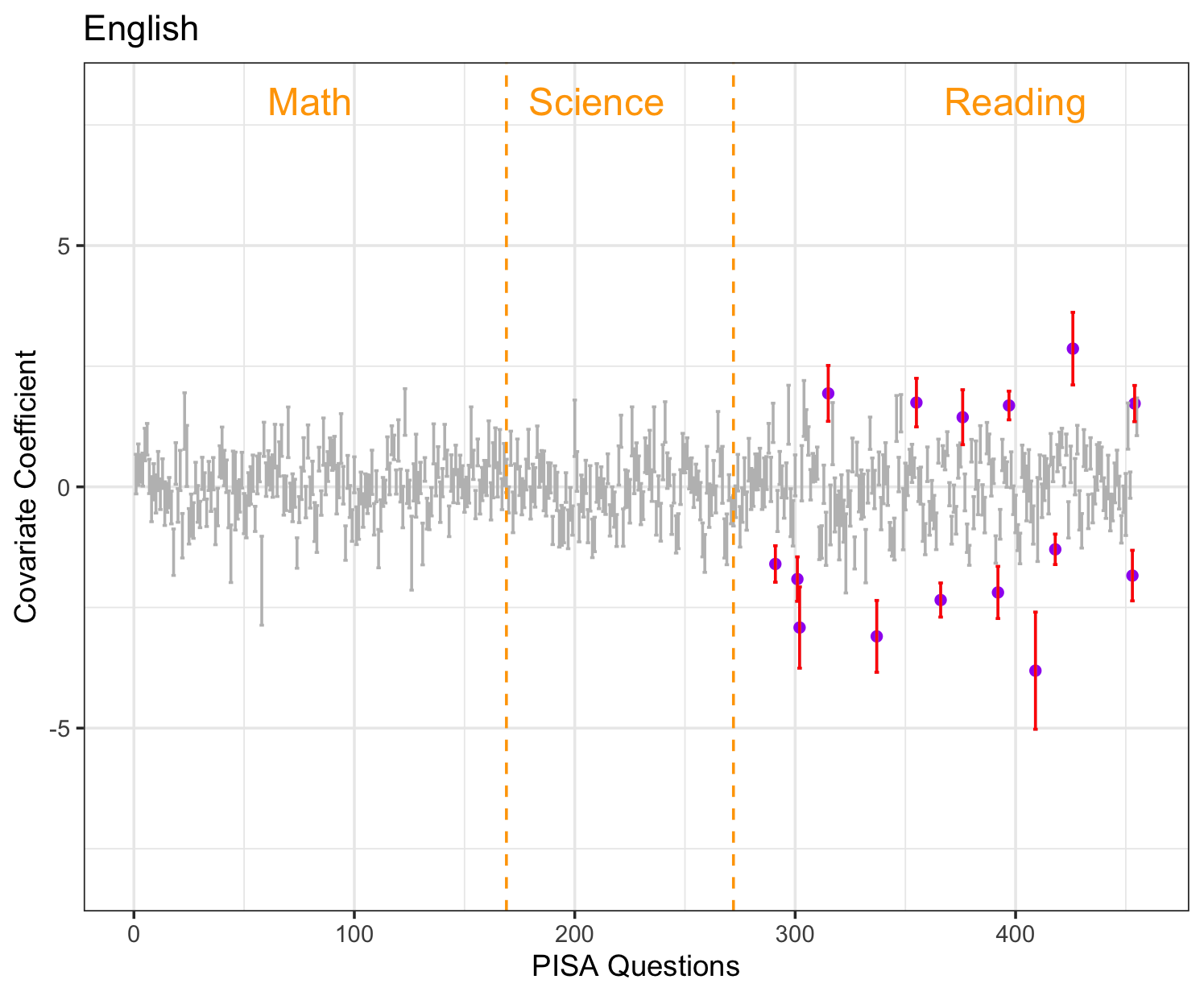}~\includegraphics[width=0.5\linewidth]{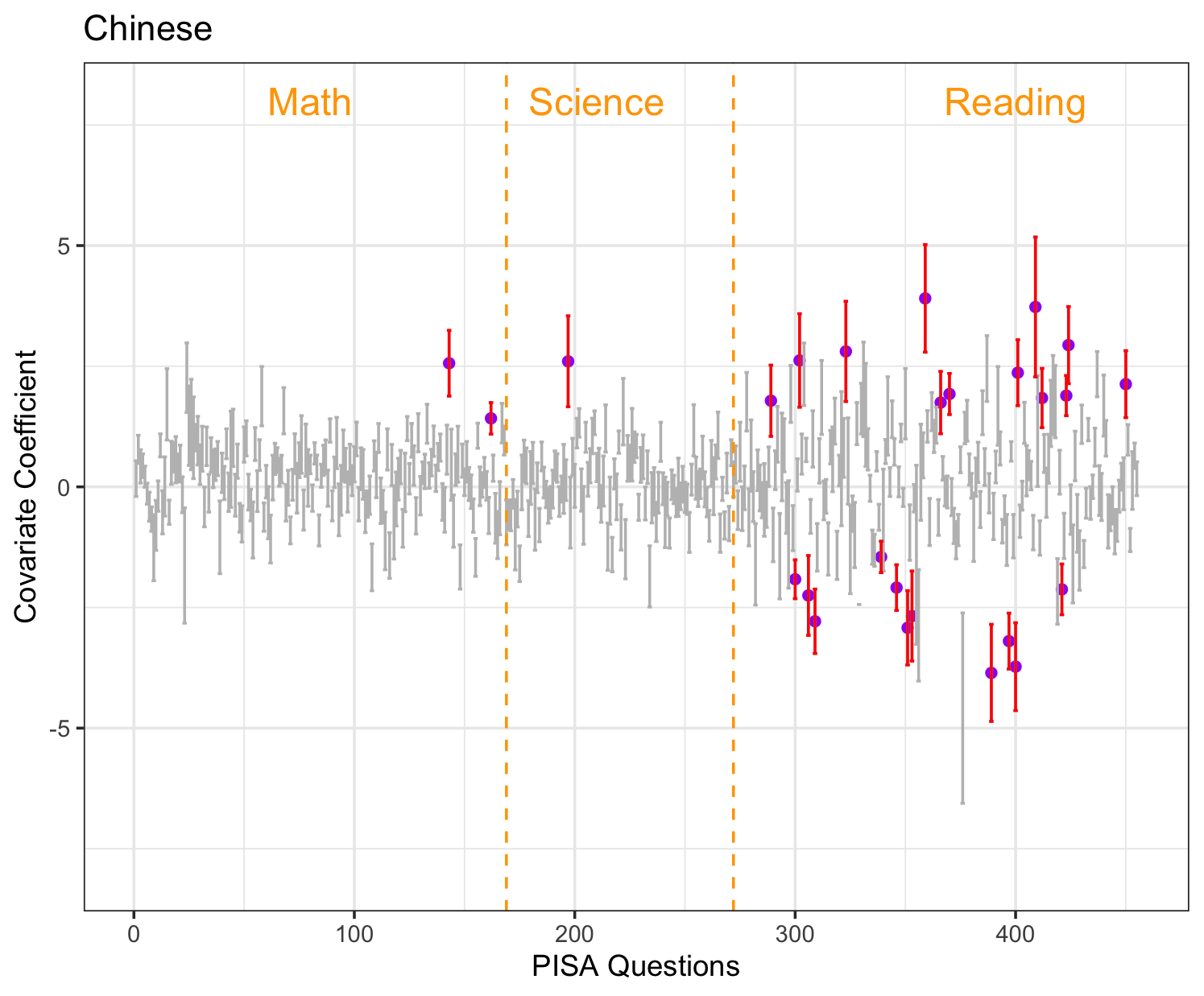}\\
    \includegraphics[width=0.5\linewidth]{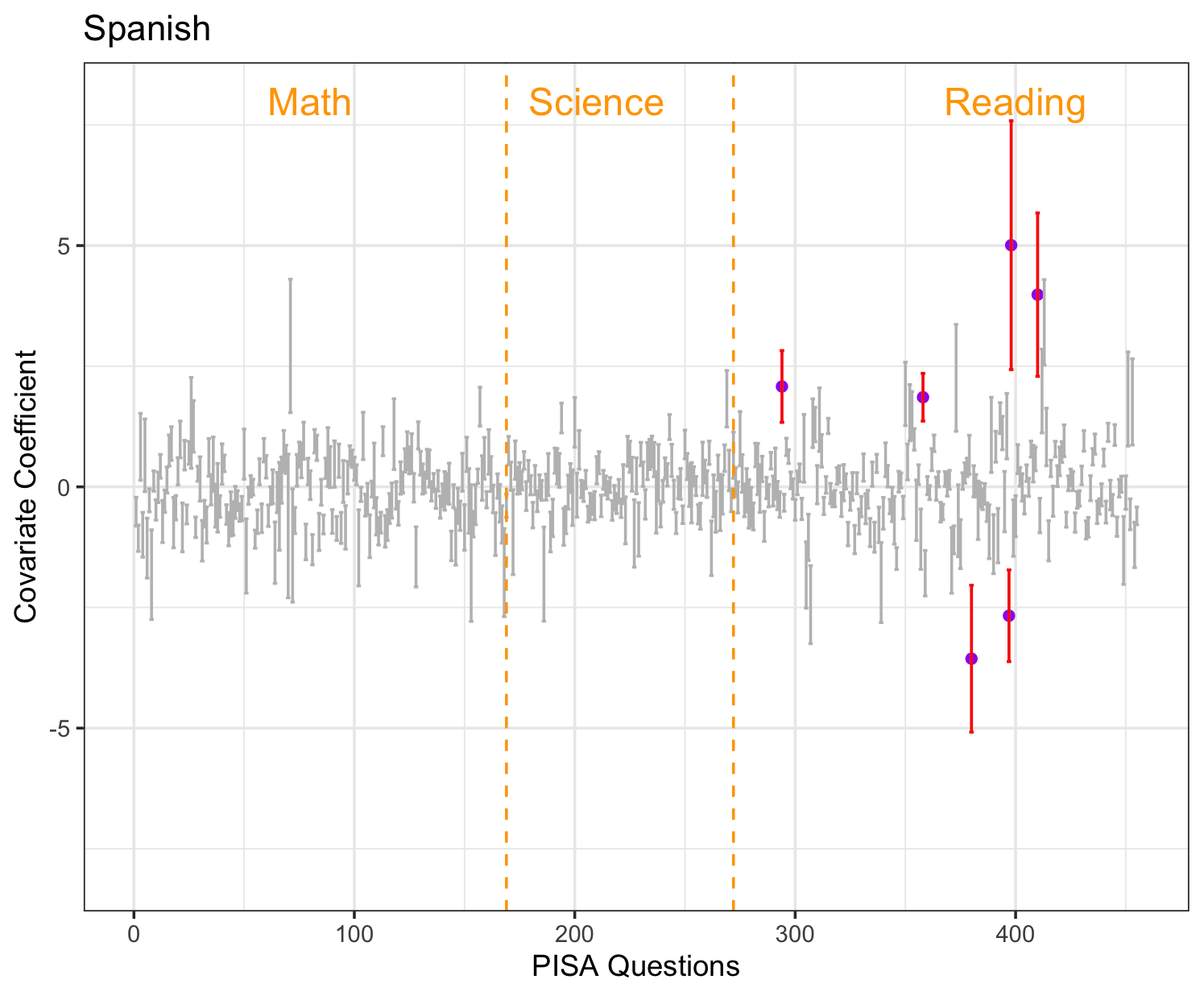}~
    \includegraphics[width=0.5\linewidth]{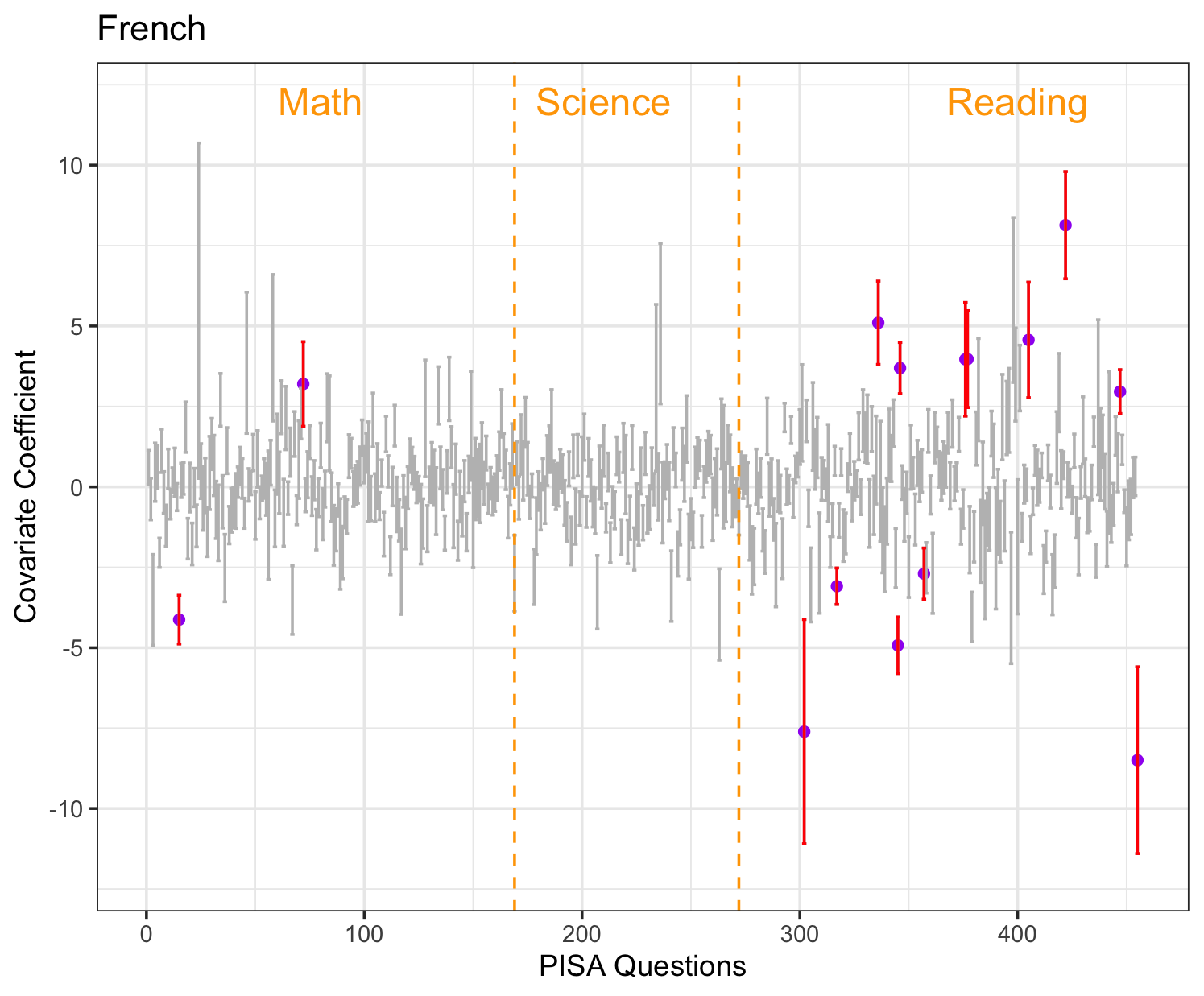}
    \caption{Confidence intervals for effects of selective assessment language indicators on each PISA item. Red intervals correspond to confidence intervals for items with significant bias after Bonferroni correction.}
    \label{fig:language bias}
\end{figure}

We apply the proposed method to the processed PISA data to estimate the covariate effect for each item and to construct the corresponding $95\%$ confidence intervals. The individual covariates include indicators for each of 80 countries/regions of origin and each of 58 assessment languages.
Figure~\ref{fig:countries} summarizes that for each country-of-origin indicator, the number of items detected as biased after Bonferroni correction. To investigate these covariate effects in detail, we focus on several countries and present the point and interval estimates for the effect of the country-of-origin indicators on each item in Figure~\ref{fig:country bias}. The analogous results for each of the assessment language indicators are presented in Figure~\ref{fig:language bias}. In Figures~\ref{fig:country bias}--\ref{fig:language bias}, intervals highlighted in red correspond to biased items with significant covariate effect after Bonferroni Correction. For comparison, we also apply a baseline method that fits an 
$L_1$-penalized GLM without adjusting for latent variables and analogous results are deferred to Section~H of Supplementary Materials due to space constraints.

For country-of-origin indicators, the top panel of Figure~\ref{fig:country bias} shows that for the USA or the United Kingdom, relatively few items are biased, and most of the biased items are in the reading domain. The bottom left panel of Figure~\ref{fig:country bias} shows that for Singapore, the number of biased items is larger than that of USA or United Kingdom, and the biased items spread across three domains, although there are still a bit more biased items in the reading domain. From the bottom right panel of Figure~\ref{fig:country bias}, we notice an evident pattern for Chinese Taipei that most significantly biased items are in the reading domain, a small number of biased items appear in the math domain, and only a single science item is detected to be biased. These patterns suggest that potential measurement bias arises more frequently in reading items.
When a significant interval is entirely above zero, test-takers from that country have a higher probability of answering the item correctly than others, conditioned on the same latent ability. Intervals entirely below zero indicate the opposite. For the United States, most detected items lie above zero, whereas for the United Kingdom, Singapore, and Chinese Taipei we observe a mix of positive and negative covariate effects across items. 
For language indicators, in Figure~\ref{fig:language bias}, we also observe a large amount of significantly biased items are in the reading domain across the four assessment languages considered. This pattern is consistent with the occurrence of translation-induced issues in question context descriptions. Mathematics and Science domains show fewer biased items, which aligns with the findings that misfit across country/language groups in PISA is typically more evident in reading than in mathematics or science~\citep{joo2022impact, oecd2024}. 
One possible explanation is that mathematics and science items often rely more heavily on symbolic representations, which may be less sensitive to translation differences \citep{mullis2011impact}.
 Where language effects are significant, we see both positive and negative intervals, indicating no consistent advantage for any single language~\citep{zhu2022investigation}. In the baseline method without accounting for latent variables, substantially more items are detected as exhibiting DIF. This likely occurs because variation due to latent abilities is absorbed by the observed covariates, which inflates the estimated covariate effects and leads to over-detection of DIF items. See Section~H of Supplementary Materials for more details.

\begin{figure}[h!]
    \centering
   \includegraphics[width=0.7\linewidth]{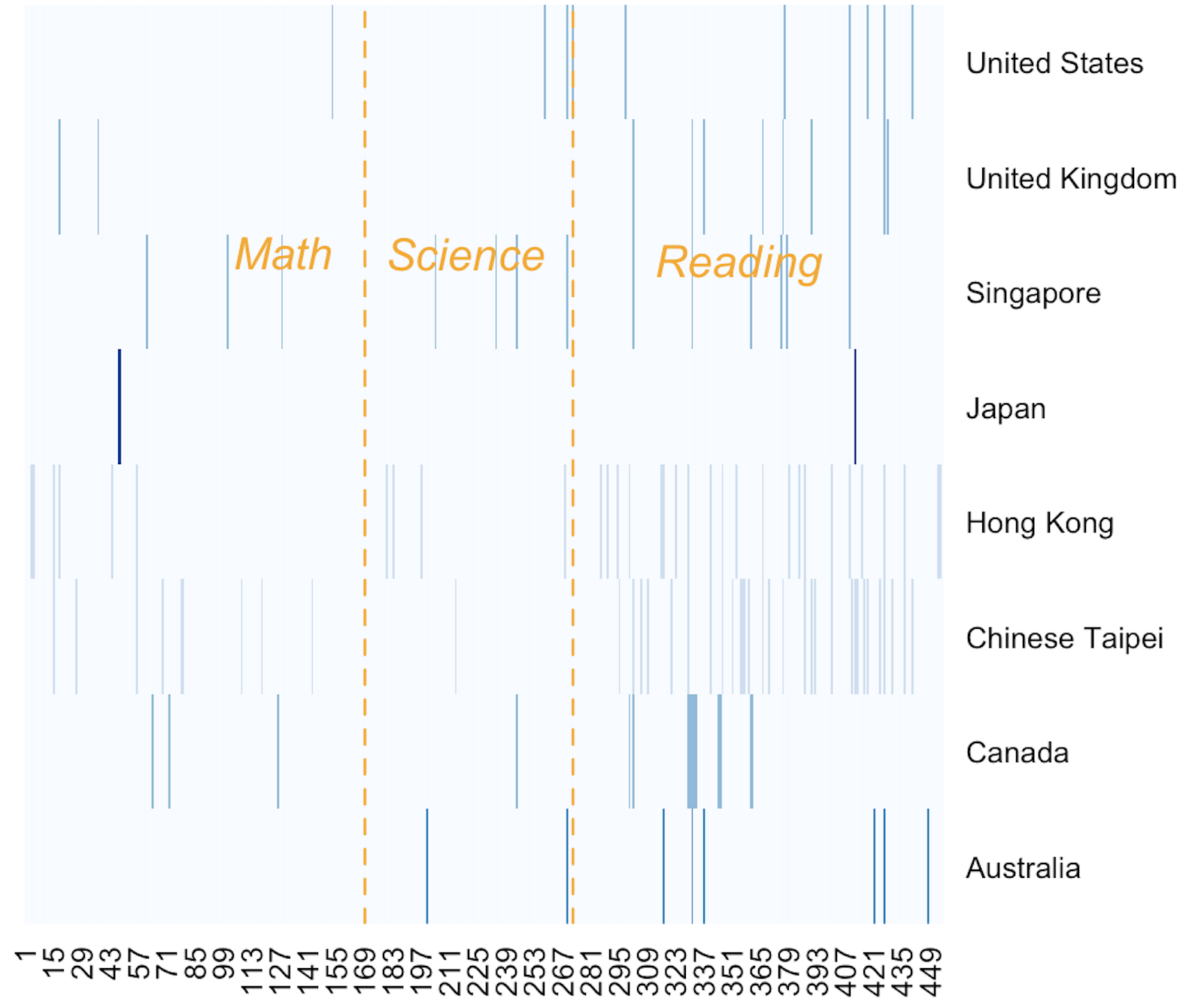}
    \caption{Heatmap plot of significantly nonzero covariate effect for selected country-of-origin indicators by item. Darker blue indicates a greater amount of biased items detected after the Bonferroni Correction. }
    \label{fig:heatmap country}
\end{figure}

\begin{figure}[h!]
    \centering
    \includegraphics[width=0.9\linewidth]{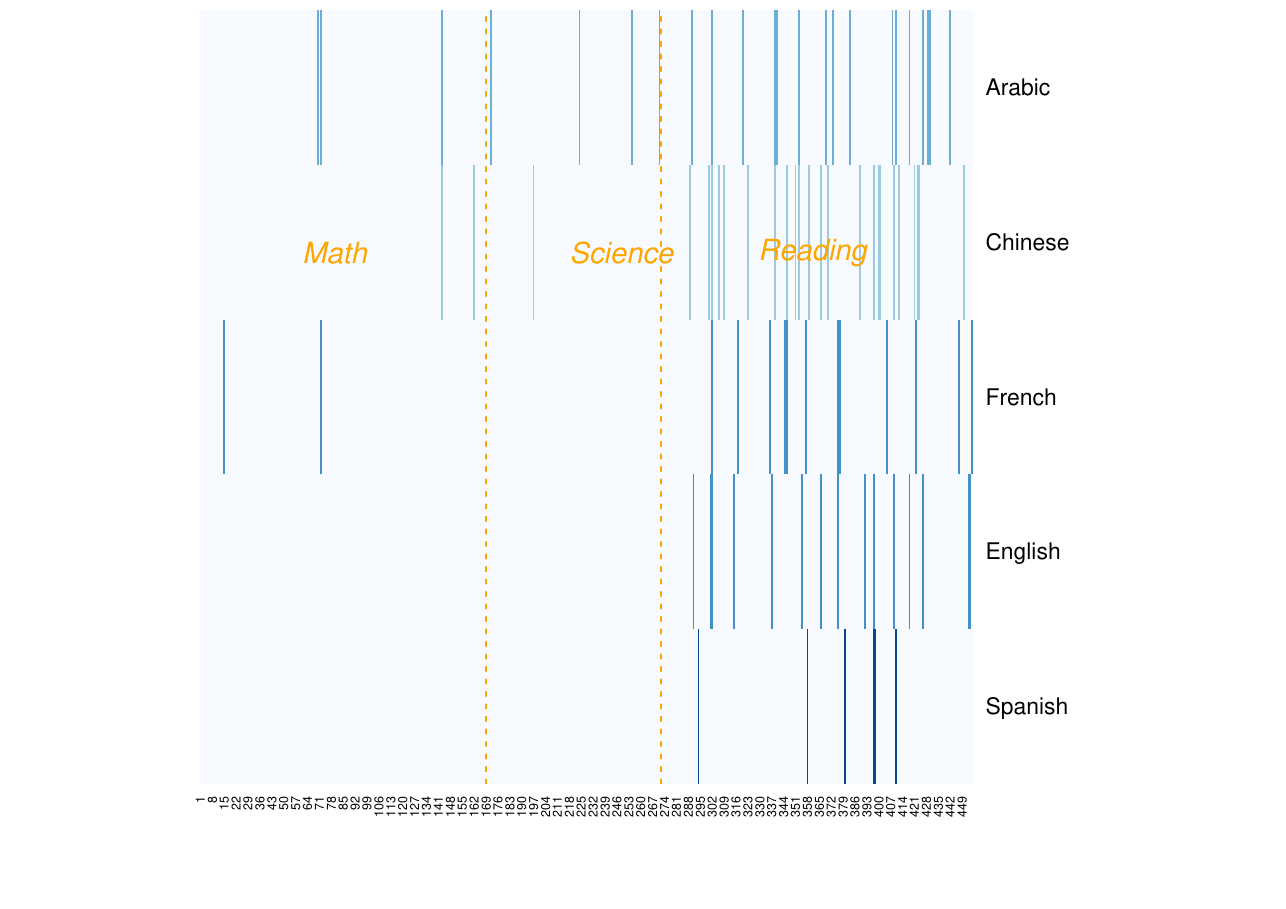}
    \caption{Heatmap plot of significantly nonzero covariate effect for selected language assessment indicators by item. Darker blue indicates a greater amount of biased items detected after the Bonferroni Correction.}
    \label{fig:heatmap language}
\end{figure}

In addition, we generate heatmaps of the identified biased items across different country groups in Figure~\ref{fig:heatmap country}, and across different languages of assessments in Figure~\ref{fig:heatmap language}. Due to the display limitation, not all of 455 items can be presented in the figure, instead, each of several adjacent items are grouped in one $x$-tick, and the color scale encodes the number of biased items among the small group of items, with darker colors indicating more biased items among these small group of items corresponding to each of country of origin indicator. We notice that
Across all selected countries, biased item clusters occur less frequently for Japan, and there is a considerable amount of biased items in the United States, the United Kingdom, Singapore, Canada, and Australia, whereas biased items occur more often in Hong Kong and Taipei. A darker color scale is observed overall in the reading domains. For the heatmap corresponding to the language of assessment, it is observed that for English and Spanish, all the biased items are in the reading domains. For the other three remaining languages: Arabic, Chinese, and French, most of the biased items are in reading domains. The Chinese language also has a dense cluster of biased items in reading domains.

The large amount of item biases that occur in reading domains is consistent with cultural or curricular differences in reading passages, translation-related tonal differences that can subtly affect difficulty, and differences in students’ exposure to particular text genres~\citep{petersen2003use, gregoire2018itc, oecd2024}. Although our estimates capture covariate effects after adjusting for latent ability and many covariates, they do not, by themselves, identify the source of bias; items detected as biased by our method should be prioritized for expert review and, where appropriate, revision or retranslation.

\section{Discussion and Future Directions}
\label{sec:discussion}

In this paper, we investigate a high-dimensional generalized latent variable model designed for discrete or mixed-type responses with high-dimensional covariates. Our model allows for flexible dependence between covariates and latent variables, a complexity that renders standard one-step regularized estimators theoretically intractable. To address this challenge, we develop an alternating algorithm for estimating the covariate effects and, based on these estimators, construct a debiased estimator for inference. We establish rigorous theoretical properties for our estimators—including sharp error bounds, estimation consistency, and asymptotic normality—providing a foundation for valid inference in this complex, high-dimensional setting.

% Specifically, conditioned on the current latent variables, we update the item-specific covariate effects and loading parameters via regularized generalized linear regressions; conditioned on the updated covariate effects and loading parameters, we then update the latent variables through generalized linear regressions. Iterating these two steps yields theoretically consistent estimators. Moreover, we also develop first-order approximate algorithms for solving these convex subproblems. Under mild conditions on the number of iterations, the resulting estimators from approximate algorithms achieve the same statistical consistency as the exact minimizers. 

One limitation of the proposed method is its reliance on a robust initial estimator. Currently, our initial estimator uses a simplified model that excludes covariates, which may be improved.
A promising direction is to incorporate the low-rank factor structure via nuclear-norm regularization, a technique widely used in the literature on matrix estimation~\citep{koltchinskii2011nuclear, koltchinskii2015optimal}. 
Such an approach may help relax the requirement in Assumption~\ref{assumption:intialization} on the initialization. 
While nuclear norm regularization transforms the estimation problem into a convex optimization problem, choosing the penalty weight requires careful theoretical analysis, which is nontrivial due to the flexible dependence between the covariates and the latent variables. Furthermore, nuclear norm regularization yields only an approximately low-rank structure; therefore, additional projection steps may be needed.

Beyond refining the initial estimator, this research offers several promising avenues for extension. First, our current analysis utilizes a fully parametric generalized latent variable model, which may be overly restrictive in complex applications. Specifically, when the goal is to isolate the effect of a primary covariate while accounting for latent variables and high-dimensional control variables, the model could be extended to include a nonparametric component—akin to the partially linear framework of Robinson (1988). Valid statistical inference on the effect of the primary covariate may be achieved by integrating our approach with doubly debiased machine learning (Chernozhukov et al., 2018). Second, extending this framework to accommodate group-wise covariate effects is of significant interest. This is particularly relevant for high-dimensional categorical predictors (e.g., evaluating the joint effect of multiple dummy variables representing geographic regions). By leveraging simultaneous inference techniques \citep{zhang2017simultaneous}, bootstrap-assisted procedures could be employed to derive group-wise asymptotic distributions, facilitating valid hypothesis testing and the construction of joint confidence regions.

 \spacingset{1.18} % DON'T change the spacing

\bibliographystyle{apalike}
\bibliography{reference}

\begin{thebibliography}{}

\bibitem[Bai, 2003]{bai2003inferential}
Bai, J. (2003).
\newblock Inferential theory for factor models of large dimensions.
\newblock {\em Econometrica}, 71(1):135--171.

\bibitem[Bai and Ng, 2002]{bai2002determining}
Bai, J. and Ng, S. (2002).
\newblock Determining the number of factors in approximate factor models.
\newblock {\em Econometrica}, 70(1):191--221.

\bibitem[Bartholomew et~al., 2011]{bartholomew2011latent}
Bartholomew, D.~J., Knott, M., and Moustaki, I. (2011).
\newblock {\em Latent variable models and factor analysis: A unified approach}.
\newblock John Wiley \& Sons.

\bibitem[Bing et~al., 2024]{bing2023inference}
Bing, X., Cheng, W., Feng, H., and Ning, Y. (2024).
\newblock Inference in high-dimensional multivariate response regression with
  hidden variables.
\newblock {\em Journal of the American Statistical Association},
  119(547):2066--2077.

\bibitem[Bing and Wegkamp, 2019]{bing2019adaptive}
Bing, X. and Wegkamp, M.~H. (2019).
\newblock Adaptive estimation of the rank of the coefficient matrix in
  high-dimensional multivariate response regression models.
\newblock {\em The Annals of Statistics}, 47(6):3157--3184.

\bibitem[B{\"u}hlmann and Van De~Geer, 2011]{buhlmann2011statistics}
B{\"u}hlmann, P. and Van De~Geer, S. (2011).
\newblock {\em Statistics for high-dimensional data: methods, theory and
  applications}.
\newblock Springer Science \& Business Media.

\bibitem[{\'C}evid et~al., 2020]{Domagoj2020}
{\'C}evid, D., B{\"u}hlmann, P., and Meinshausen, N. (2020).
\newblock Spectral deconfounding via perturbed sparse linear models.
\newblock {\em Journal of Machine Learning Research}, 21(232):1--41.

\bibitem[Chen et~al., 2023]{chen2023dif}
Chen, Y., Li, C., Ouyang, J., and Xu, G. (2023).
\newblock {DIF} statistical inference without knowing anchoring items.
\newblock {\em Psychometrika}, 88(4):1097--1122.

\bibitem[Chen and Li, 2022]{chen2022determining}
Chen, Y. and Li, X. (2022).
\newblock Determining the number of factors in high-dimensional generalized
  latent factor models.
\newblock {\em Biometrika}, 109(3):769--782.

\bibitem[Dobriban, 2020]{dobriban2020permutation}
Dobriban, E. (2020).
\newblock Permutation methods for factor analysis and {PCA}.
\newblock {\em The Annals of Statistics}, 48(5):2824--2847.

\bibitem[Du et~al., 2025]{du2025simultaneous}
Du, J.-H., Wasserman, L., and Roeder, K. (2025).
\newblock Simultaneous inference for generalized linear models with unmeasured
  confounders.
\newblock {\em Journal of the American Statistical Association},
  120(551):1945--1959.

\bibitem[Fan et~al., 2024]{fan2024latent}
Fan, J., Lou, Z., and Yu, M. (2024).
\newblock Are latent factor regression and sparse regression adequate?
\newblock {\em Journal of the American Statistical Association},
  119(546):1076--1088.

\bibitem[Gagnon-Bartsch and Speed, 2012]{gagnon2012using}
Gagnon-Bartsch, J.~A. and Speed, T.~P. (2012).
\newblock Using control genes to correct for unwanted variation in microarray
  data.
\newblock {\em Biostatistics}, 13(3):539--552.

\bibitem[Goplerud et~al., 2025]{goplerud2025partially}
Goplerud, M., Papaspiliopoulos, O., and Zanella, G. (2025).
\newblock Partially factorized variational inference for high-dimensional mixed
  models.
\newblock {\em Biometrika}, 112(2):asae067.

\bibitem[Gregoire, 2018]{gregoire2018itc}
Gregoire, J. (2018).
\newblock {ITC} guidelines for translating and adapting tests.
\newblock {\em International Journal of Testing}, 18(2):101--134.

\bibitem[Guo et~al., 2022]{guo2021doubly}
Guo, Z., {\'C}evid, D., and B{\"u}hlmann, P. (2022).
\newblock {Doubly debiased lasso: High-dimensional inference under hidden
  confounding}.
\newblock {\em The Annals of Statistics}, 50(3):1320--1347.

\bibitem[Holland and Wainer, 2012]{holland2012differential}
Holland, P.~W. and Wainer, H. (2012).
\newblock {\em Differential item functioning}.
\newblock Routledge.

\bibitem[Javanmard and Montanari, 2014]{javanmard2014confidence}
Javanmard, A. and Montanari, A. (2014).
\newblock Confidence intervals and hypothesis testing for high-dimensional
  regression.
\newblock {\em Journal of Machine Learning Research}, 15(1):2869--2909.

\bibitem[Joo et~al., 2022]{joo2022impact}
Joo, S., Ali, U., Robin, F., and Shin, H.~J. (2022).
\newblock Impact of differential item functioning on group score reporting in
  the context of large-scale assessments.
\newblock {\em Large-Scale Assessments in Education}, 10(18):1--21.

\bibitem[Koltchinskii et~al., 2011]{koltchinskii2011nuclear}
Koltchinskii, V., Lounici, K., and Tsybakov, A.~B. (2011).
\newblock Nuclear-norm penalization and optimal rates for noisy low-rank matrix
  completion.
\newblock {\em The Annals of Statistics}, 39(5):2302–2329.

\bibitem[Koltchinskii and Xia, 2015]{koltchinskii2015optimal}
Koltchinskii, V. and Xia, D. (2015).
\newblock Optimal estimation of low rank density matrices.
\newblock {\em Journal of Machine Learning Research}, 16(53):1757--1792.

\bibitem[Lee and Ning, 2025]{lee2025ghive}
Lee, I. and Ning, Y. (2025).
\newblock {G-HIVE: parameter estimation and approximate inference for
  multivariate response generalized linear models with hidden variables}.
\newblock {\em arXiv preprint arXiv:2509.00196}.

\bibitem[Lee et~al., 2017]{lee2017improved}
Lee, S., Sun, W., Wright, F.~A., and Zou, F. (2017).
\newblock An improved and explicit surrogate variable analysis procedure by
  coefficient adjustment.
\newblock {\em Biometrika}, 104(2):303--316.

\bibitem[Leek and Storey, 2008]{leek2008general}
Leek, J.~T. and Storey, J.~D. (2008).
\newblock A general framework for multiple testing dependence.
\newblock {\em Proceedings of the National Academy of Sciences},
  105(48):18718--18723.

\bibitem[Loh and Wainwright, 2015]{loh2013regularized}
Loh, P.-L. and Wainwright, M.~J. (2015).
\newblock Regularized {M}-estimators with nonconvexity: Statistical and
  algorithmic theory for local optima.
\newblock {\em Journal of Machine Learning Research}, 16(19):559--616.

\bibitem[McKennan and Nicolae, 2019]{mckennan2019accounting}
McKennan, C. and Nicolae, D. (2019).
\newblock Accounting for unobserved covariates with varying degrees of
  estimability in high-dimensional biological data.
\newblock {\em Biometrika}, 106(4):823--840.

\bibitem[Mullis et~al., 2011]{mullis2011impact}
Mullis, I.~V., Martin, M.~O., and Foy, P. (2011).
\newblock The impact of reading ability on timss mathematics and science
  achievement at the fourth grade: An analysis by item reading demands.
\newblock {\em TIMSS and PIRLS}, pages 67--108.

\bibitem[Newey and McFadden, 1994]{newey1994large}
Newey, W.~K. and McFadden, D. (1994).
\newblock Large sample estimation and hypothesis testing.
\newblock {\em Handbook of Econometrics}, 4:2111--2245.

\bibitem[Ning and Liu, 2017]{ning2017}
Ning, Y. and Liu, H. (2017).
\newblock {A general theory of hypothesis tests and confidence regions for
  sparse high dimensional models}.
\newblock {\em The Annals of Statistics}, 45(1):158--195.

\bibitem[OECD, 2024]{oecd2024}
OECD (2024).
\newblock {PISA} 2022 technical report.
\newblock {\em PISA, OECD Publishing, Paris}.

\bibitem[Ouyang et~al., 2026]{ouyang2025statistical}
Ouyang, J., Cui, C., Tan, K.~M., and Xu, G. (2026).
\newblock Statistical inference for covariate-adjusted and interpretable
  generalized latent factor model with application to testing fairness.
\newblock {\em The Annals of Applied Statistics}, 20(1):764--788.

\bibitem[Ouyang et~al., 2023]{ouyang2023high}
Ouyang, J., Tan, K.~M., and Xu, G. (2023).
\newblock High-dimensional inference for generalized linear models with hidden
  confounding.
\newblock {\em Journal of Machine Learning Research}, 24(296):1--61.

\bibitem[Pandolfi et~al., 2025]{pandolfi2024conjugate}
Pandolfi, A., Papaspiliopoulos, O., and Zanella, G. (2025).
\newblock Conjugate gradient methods for high-dimensional {GLMMs}.
\newblock {\em Journal of the American Statistical Association, in press}.

\bibitem[Petersen et~al., 2003]{petersen2003use}
Petersen, M.~A., Groenvold, M., Bjorner, J.~B., Aaronson, N., Conroy, T., Cull,
  A., Fayers, P., Hjermstad, M., Sprangers, M., and Sullivan, M. (2003).
\newblock Use of differential item functioning analysis to assess the
  equivalence of translations of a questionnaire.
\newblock {\em Quality of Life Research}, 12(4):373--385.

\bibitem[Schleicher, 2019]{schleicher2019pisa}
Schleicher, A. (2019).
\newblock {PISA} 2018: Insights and interpretations.
\newblock {\em OECD Publishing}.

\bibitem[Skrondal and Rabe-Hesketh, 2004]{skrondal2004generalized}
Skrondal, A. and Rabe-Hesketh, S. (2004).
\newblock {\em Generalized latent variable modeling: Multilevel, longitudinal,
  and structural equation models}.
\newblock Chapman and Hall/CRC.

\bibitem[van~de Geer et~al., 2014]{van2014}
van~de Geer, S., B{\"u}hlmann, P., Ritov, Y., and Dezeure, R. (2014).
\newblock {On asymptotically optimal confidence regions and tests for
  high-dimensional models}.
\newblock {\em The Annals of Statistics}, 42(3):1166--1202.

\bibitem[Wainwright, 2019]{wainwright2019high}
Wainwright, M.~J. (2019).
\newblock {\em High-Dimensional Statistics: A Non-Asymptotic Viewpoint}.
\newblock Cambridge University Press.

\bibitem[Wang, 2022]{wang2022maximum}
Wang, F. (2022).
\newblock Maximum likelihood estimation and inference for high dimensional
  generalized factor models with application to factor-augmented regressions.
\newblock {\em Journal of Econometrics}, 229(1):180--200.

\bibitem[Wang et~al., 2017]{wang2017confounder}
Wang, J., Zhao, Q., Hastie, T., and Owen, A.~B. (2017).
\newblock Confounder adjustment in multiple hypothesis testing.
\newblock {\em The Annals of statistics}, 45(5):1863.

\bibitem[Wang and Shah, 2025]{wang2025latent}
Wang, Y. and Shah, R. (2025).
\newblock Latent confounding in high-dimensional nonlinear models.
\newblock {\em arXiv preprint arXiv:2508.06274}.

\bibitem[Xia et~al., 2018]{xia2018joint}
Xia, Y., Cai, T.~T., and Li, H. (2018).
\newblock Joint testing and false discovery rate control in high-dimensional
  multivariate regression.
\newblock {\em Biometrika}, 105(2):249--269.

\bibitem[Zhang and Zhang, 2014]{zhang2014}
Zhang, C.-H. and Zhang, S.~S. (2014).
\newblock Confidence intervals for low dimensional parameters in high
  dimensional linear models.
\newblock {\em {Journal of the Royal Statistical Society: Series B
  (Methodological)}}, 76(1):217--242.

\bibitem[Zhang et~al., 2020]{zhang2020note}
Zhang, H., Chen, Y., and Li, X. (2020).
\newblock A note on exploratory item factor analysis by singular value
  decomposition.
\newblock {\em Psychometrika}, 85(2):358--372.

\bibitem[Zhang and Cheng, 2017]{zhang2017simultaneous}
Zhang, X. and Cheng, G. (2017).
\newblock Simultaneous inference for high-dimensional linear models.
\newblock {\em Journal of the American Statistical Association},
  112(518):757--768.

\bibitem[Zhu and Aryadoust, 2022]{zhu2022investigation}
Zhu, X. and Aryadoust, V. (2022).
\newblock An investigation of mother tongue differential item functioning in a
  high-stakes computerized academic reading test.
\newblock {\em Computer Assisted Language Learning}, 35(3):412--436.

\end{thebibliography}

\end{document}